\documentclass[lettersize,journal]{IEEEtran}
\usepackage{amsmath,amssymb,amsfonts}
\usepackage{algorithm,algorithmic}
\usepackage{array}
\usepackage{cite}
\usepackage{textcomp}
\usepackage{stfloats}
\usepackage{setspace}
\usepackage{url}
\usepackage{xcolor}
\usepackage{multirow}
\usepackage{verbatim}
\usepackage{graphicx}
\usepackage{diagbox}
\usepackage{subfig}
\usepackage{hyperref}
\usepackage{bm}
\allowdisplaybreaks[4]
\hypersetup{
	colorlinks=true,
	linkcolor=blue,
	filecolor=blue,
	urlcolor=blue,
	citecolor=cyan,
}
\usepackage{balance}
\def \blue {\textcolor[rgb]{0.00,0.00,0.00}}

\begin{document}
	\title{Joint Positioning, Beamforming, and Power Allocation in Full-Duplex MIMO with Position-Reconfigurable Antenna Arrays}
	\author{Chengjie Zhao, Yuanzhe Gong, Tho Le-Ngoc, ~\IEEEmembership{Life Fellow, IEEE}
		\thanks{Chengjie Zhao, Yuanzhe Gong, and Tho Le-Ngoc are with the Department of Electrical and Computer Engineering, McGill University, Montreal, QC H3A 0E9, Canada (e-mail: chengjie\_zhao1214@163.com; yuanzhe.gong@mail.mcgill.ca; tho.le-ngoc@mcgill.ca).}}
	
	\maketitle

	\begin{abstract}
		We consider a multi-user (MU) full-duplex (FD) multiple-input multiple-output (MIMO) communication system, in which the base station transceiver is equipped with  transmit and receive position reconfigurable antennas (PRAs) to mitigate both MU-interference (MUI) and self-interference (SI) while enhancing the desired signal quality. We first formulate a joint design of beamforming, transmit power allocation, and antenna placement to maximize the weighted sum-rate of the considered PRA-based MU-FD-MIMO system under the constraints on reconfigurable region size, minimum antenna spacing, and transmit power. To address this highly non-convex problem, we then propose an alternating optimization (AO) framework that decomposes the original problem into subproblems and solves them iteratively. In particular, fractional programming techniques are used to separate the optimization variables from the logarithmic and ratio terms, while the block successive upper bound minimization method handles the non-convexity of PRA placement. Simulation results confirm a significant performance gain when integrating PRAs into the MU-FD-MIMO system and demonstrate the advantage of the proposed optimization framework. 
	\end{abstract}
	
	\section{Introduction}
	Wireless service providers operate in a rapidly evolving environment defined by ever-growing demands for higher data rates, enhanced energy and spectrum efficiency, expanded capacity, and greater link reliability. Multi-antenna systems have become a cornerstone technology for meeting these challenges. By exploiting spatial multiplexing and diversity, antenna arrays can transmit and receive multiple data streams simultaneously, dramatically increasing spectral efficiency without additional bandwidth or power consumption. Moreover, optimized beamforming enables coherent enhancement of desired signals while precisely suppressing interference, delivering substantial improvements in signal-to-interference-plus-noise ratio (SINR) and system throughput \cite{MIMO}. \\
	\indent Furthermore, to accommodate the explosive growth of smart devices, conventional half‐duplex base‐station operation, via frequency‐division duplexing (FDD) or time‐division duplexing (TDD), faces inherent limitations in spectral efficiency and latency. In high‐density scenarios, FDD places a heavy burden on scarce spectrum resources, while TDD's reliance on strict time‐slot orthogonality can introduce latency and synchronization challenges that hinder timely interaction among many connected devices \cite{IoT}. To break this bottleneck, in‐band full‐duplex (FD) enables simultaneous transmission and reception over the same time–frequency resource, offering the potential to double spectral efficiency and substantially reduce latency compared to traditional HD systems \cite{FD, FD1}. This capability makes FD a compelling solution to alleviate spectrum scarcity and enhance network efficiency in advanced wireless communication systems. However, a fundamental challenge in FD communication is the self-interference (SI), which arises from the co-located transmitter and receiver at the same node \cite{FD}. \\
	\indent The employment of antenna arrays at BS can provide rich spatial multiplexing gain to direct signals to users of interest while suppressing the SI and substantial array gain to compensate for severe propagation losses and mitigate interference \cite{mMIMO}. Nonetheless, in traditional systems, antennas are typically arranged in fixed geometries, spaced half a wavelength apart, and once manufactured, they can only passively adapt to channel conditions by adjusting beamformer to meet communication requirements. Consequently, conventional approaches for handling SI  usually lead to a trade-off between SI suppression and desired signal preservation \cite{FD1}. For example in \cite{FD3}, the desired beam is tilted to align the radiation nulls with the strong SI direction. \\	
	\indent Alternatively, multiple-input multiple-output (MIMO) systems incorporating position-reconfigurable antennas (PRA) provide a promising solution to enhance both user-directed transmission and SI suppression efficiency. PRA systems (PRAS) have recently emerged and been implemented in various ways like flexible antenna systems (FAS) \cite{FAS}, and movable antenna systems (MAS) \cite{MAS}. PRAS dynamically reconfigures antenna positions and inter-element geometry to enable controllable electromagnetic interactions. The antenna positioning essentially allows the system to move from areas with poor channel conditions to more favorable locations, thereby enhancing signal quality \cite{FAS, MAS}. When combined with strategic beamformer design, this approach more efficiently explores spatial channel variations. By controlling both self-interference and transmit channels, PRAS fully exploits an antenna array’s spatial diversity, generating richer beam patterns \cite{MASFAG, MASMBF, MAS-FBC} and enabling high-quality signal transmission with simultaneous SI suppression. \\
	\indent With its ability to adjust the antenna position and actively adapt to the environment, single-antenna PRAS has demonstrated performance comparable to that of a single-input multiple-output (SIMO) beamforming system \cite{MAS2, FAS2}. Furthermore, PRAS-assisted MIMO systems have been shown to significantly enhance MIMO channel capacity \cite{MIMOMA}, thereby improving overall communication quality. By leveraging additional degrees of freedom (DoF) in the spatial domain, PRAS presents a promising approach to enhance communication performance in terms of both achievable data rates and hardware efficiency. \\		
	\indent With these advantages, PRAS is regarded as a potential enabler of next-generation wireless communication technologies and is envisioned to have promising applications in future Internet of Things (IoT) networks, such as industrial IoT, smart homes, and robotic networks \cite{MAS, MAS1}. Therefore, integrating PRA with FD communications is expected to further improve communication performance, making it a promising approach for next-generation wireless networks. \\
	\indent \blue{To address the inherent problem in FD systems, this paper considers the use of PRA in the conventional multi-user (MU) FD system and investigates the potential enhancement that PRAS can bring to MU-FD-MIMO systems through a joint optimization of antenna positioning, beamforming, and power allocation based on a weighted sum-rate metric.} Existing research has explored various strategies to determine the optimal antenna positions and corresponding beamforming matrices in various PRAS-assisted communication applications \cite{MIMOMA, MASFAG, MASMBF, MAS-FBC, MAS-SS, MAS-SFD-MU, MAS-FD-Sat, MAS-BCD, MAS-CCFD}. Since the general optimization problem involves multiple interdependent variables and highly non-convex objective functions, the alternating optimization (AO) framework has emerged as a dominant paradigm for efficiently solving such complex optimization tasks. \\
	\indent Specifically, for beamforming design, successive convex approximation (SCA) is a widely adopted approach for addressing complex and non-convex objective. By iteratively solving a convex surrogate of the original objective function, SCA can lead to high-quality solutions. In \cite{MAS-SS}, an MAS-assisted cognitive radio system is studied, where beamforming optimization is achieved through first-order Taylor expansion-aided SCA. Similar techniques have been applied to multi-beamforming, physical layer security in FD MIMO and satellite communication systems \cite{MASMBF, MAS-SFD-MU, MAS-FD-Sat}. Additionally, block coordinate descent and semi-definite relaxation have proven to be effective optimization methods, depending on the specific problem structure \cite{MAS-FBC}, \cite{MAS-BCD}. \\
	\indent For antenna position optimization, in addition to exhaustive search methods \cite{MAS-SS} and gradient descent (GD) \cite{FPGD}, nature-inspired particle swarm optimization (PSO) is widely utilized due to its strong capability in handling non-convex problems \cite{MAS-SFD-MU, MAS-FD-Sat, MAS-CCFD}. However, the performance of PSO is highly dependent on initialization, and it often requires significant computational and time resources to achieve convergence. To address these challenges, SCA has also been explored for antenna position optimization \cite{MASMBF, MAS-FBC}. Beyond conventional optimization techniques, machine learning-based approaches have been introduced to jointly optimize channel estimation, beamforming, and port selection, as demonstrated in \cite{CE-FAS-AI}. \\
	\indent Many existing studies have explored the integration of PRA into FD systems. In \cite{MAS-CCFD}, antenna position optimization is investigated for a double-side FD MAS with a single antenna. While the proposed approach can be extended to multi-antenna scenarios, strategic beamformer design is crucial to fully harness the benefits of multi-antenna systems. Additionally, physical layer security and power saving in \blue{MU} and satellite FD MAS have also been examined in \cite{MAS-FD-Sat} and \cite{MAS-SFD-MU}. However, to the best of our knowledge, a comprehensive analysis of the performance gains that PRAS and FD can jointly provide over conventional fixed-position antenna systems (FPAS), particularly when incorporating joint optimal beamforming, power allocation, and antenna position optimization, has yet to be conducted.	Furthermore, although the optimization techniques employed in these studies, primarily PSO and SCA, are effective for addressing non-convex problems, their heuristic nature may overlook the underlying structural properties of the optimization problem, leading to unpredictable convergence behavior. \\ 
	\indent Motivated by the above observations, in this paper, we study the joint optimization of beamforming, transmit power allocation, and antenna positions to maximize the weighted sum of achievable rates for a PRAS-aided MU-FD-MIMO system. \blue{The main contributions lie in the system-level novelty and are summarized as follows:}
	\begin{itemize}
		\item We first develop a mathematical model for a MU-FD-MIMO communication system, where the BS is equipped with separate transmit and receive arrays, each consisting of multiple PRAs. The transmitter and receiver simultaneously serve multiple down-link (DL) and up-link (UL) users in presence of both MU-interference (MUI) and SI. Based on this model, we formulate an optimization problem to jointly design the beamforming matrices and antenna positions at both the BS transmitter and receiver, and the transmission power for each DL user. The objective is to maximize the weighted sum of DL and UL achievable rates, subject to constraints on the adjustable region, power budget, and minimum inter-antenna spacing.
		\item To solve the formulated problem, we propose an AO framework that iteratively updates the designed variables. Within this framework, fractional programming (FP) is employed to decompose the complex objective function. Additionally, a block successive upper-bound minimization (BSUM) technique is applied to handle the non-convex objective function with respect to antenna positions. To further address the non-convex constraints on antenna positioning, we develop an algorithm leveraging the geometric properties of the feasible set to determine optimal antenna placement. Moreover, a simplification is also introduced to this algorithm to further enhance its computational efficiency. Using FP and BSUM, the proposed algorithm would achieve faster convergence than SCA and PSO.
		\item Finally, we provide extensive simulation results to demonstrate the performance advantages brought by exploiting additional antenna DoF and FD operation in terms of both complexity and achievable rate, over the conventional FPAS in HD transmission and other existing baseline approaches. In particular, the results highlight a significant improvement in weighted achievable rate obtained by integrating PRAs with MU-FD-MIMO. Moreover, the proposed algorithm accelerates the convergence of the optimization process. It is also confirmed that the introduced simplification significantly reduces the required computational resources at a cost of only a marginal performance loss. 
	\end{itemize}
	
	\section{System Model and Problem Formulation}
	\begin{figure}[t]
		\centering
		\includegraphics[scale=0.23]{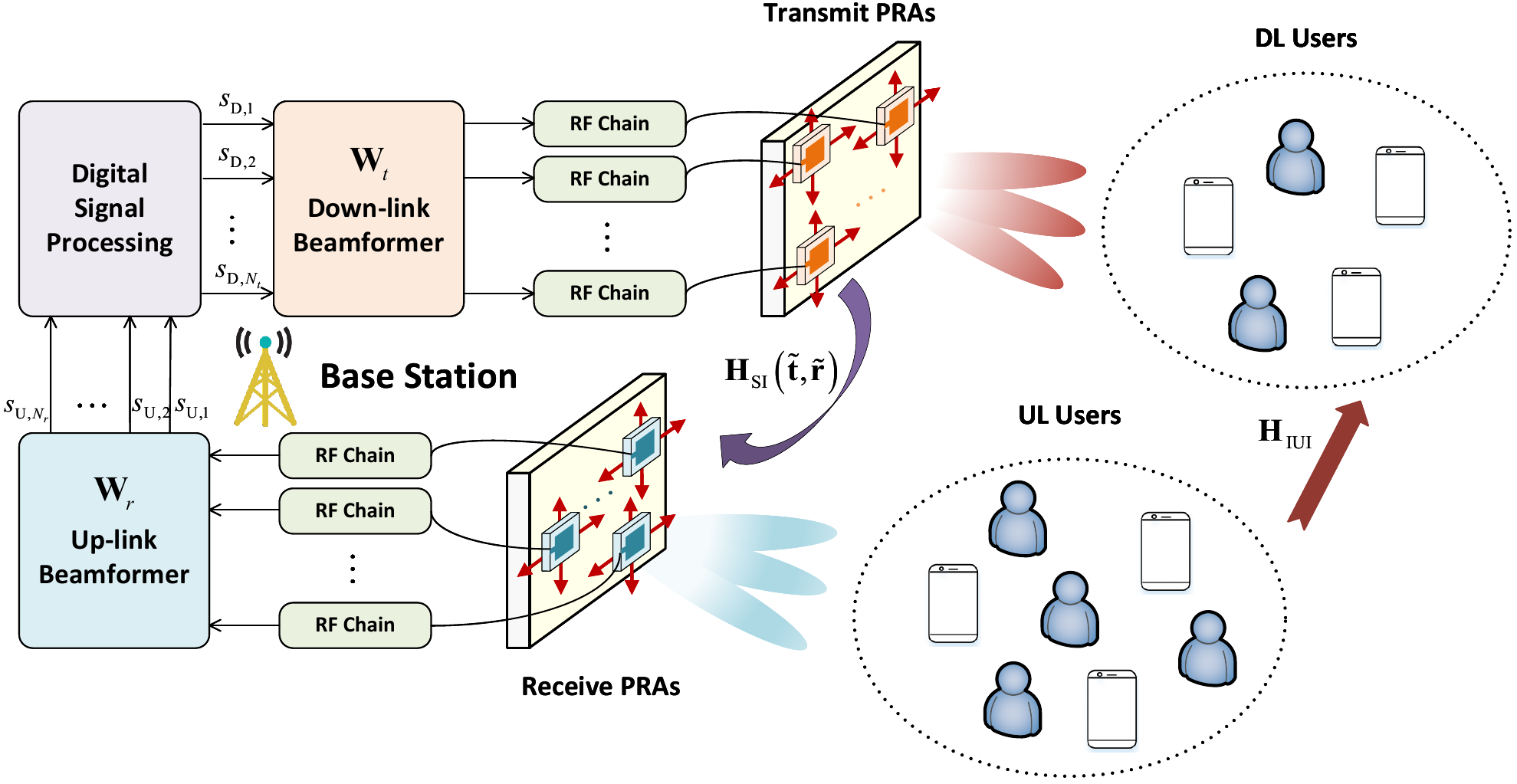}
		\caption{PRAS-aided MU-FD-MIMO communication system}	
		\label{fig1}
		\vspace{-0.5cm}
	\end{figure}
	
	\subsection{System Model}
	\indent Fig.\ref{fig1} illustrates the considered PRA-based MU-FD-MIMO communication system, in which the BS is equipped with a transmitter and receiver concurrently serving ${{K_{\rm{U}}}}$ UL and ${{K_{\rm{D}}}}$ DL users, respectively. Specifically, $N_t$ and $N_r$ PRAs are deployed at the transmitter and receiver and each PRA element is connected to the central signal processing module via a radio-frequency (RF) chain, which controls both the position and amplitude-phase adjustment. We also assume that DL and UL users have only one fixed-position antenna (FPA). \\
	\indent We define the position of the $n_t$-th (${n_t} \in {{\cal N}_t} \buildrel \Delta \over = \left\{ {n\left| {1 \le n \le {N_t},n \in \mathbb{Z}} \right.} \right\}$) transmit antenna using a Cartesian coordinate $\mathbf{t}_{n_t} = [x_{t,n_t}, y_{t,n_t}]^T \in {{\cal C}_t}$, where $\mathbb{Z}$ is the set of integer and ${\cal C}_t$ is the confined region in which the transmit PRAs can adjust their positions. Similarly, the position of the $n_r$-th (${n_r} \in {{\cal N}_r} \buildrel \Delta \over = \left\{ {n\left| {1 \le n \le {N_r},n \in \mathbb{Z}} \right.} \right\}$) receive PRA is expressed as $\mathbf{r}_{n_r} = [x_{r,n_r}, y_{r,n_r}]^T \in {{\cal C}_r}$. Without loss of generality, we assume that ${\cal C}_t$ and ${\cal C}_r$ are two-dimensional square regions with a size of $A \lambda \times A \lambda$, where $\lambda$ denotes the wavelength at the operating frequency. Furthermore, we denote the collections of transmit and receive PRA coordinates by ${\bf{\tilde t}} = \left[ {{{\bf{t}}_1},{{\bf{t}}_2}, \cdots ,{{\bf{t}}_{{N_t}}}} \right] \in {\mathbb{R}^{2 \times {N_t}}}$ and ${\bf{\tilde r}} = \left[ {{{\bf{r}}_1},{{\bf{r}}_2}, \cdots ,{{\bf{r}}_{{N_t}}}} \right] \in {\mathbb{R}^{2 \times {N_r}}}$, respectively. Then, the DL channel between the BS and the ${{k_{\rm{D}}}}$-th DL user (${{k_{\rm{D}}}} \in {{\cal K}_{\rm{D}}} \buildrel \Delta \over = \left\{ {k\left| {1 \le k \le {K_{\rm{D}}},k \in \mathbb{Z}} \right.} \right\}$) is represented as ${\bf{h}}_{{\rm{D}},{k_{\rm{D}}}}{\left( {{\bf{\tilde t}}} \right)} \in \mathbb{C}^{{N_t} \times 1}$, which depends on the transmit PRA positions, and the collection of DL channels is ${{\bf{H}}_{\rm{D}}}{\left( {{\bf{\tilde t}}} \right)} = \left[ {{{\bf{h}}_{{\rm{D}},1}}{\left( {{\bf{\tilde t}}} \right)}, \cdots ,{{\bf{h}}_{{\rm{D}},{K_{\rm{D}}}}}{\left( {{\bf{\tilde t}}} \right)}} \right] \in {\mathbb{C}^{{N_t} \times {K_{\rm{D}}}}}$. Similarly, the ${{k_{\rm{U}}}}$-th UL channel and the collection are given by ${\bf{h}}_{{\rm{U}},{k_{\rm{U}}}}{\left( {{\bf{\tilde r}}} \right)} \in \mathbb{C}^{{N_r} \times 1}$ and ${{\bf{H}}_{\rm{U}}}{\left( {{\bf{\tilde r}}} \right)} = \left[ {{{\bf{h}}_{{\rm{U}},1}}{\left( {{\bf{\tilde r}}} \right)}, \cdots ,{{\bf{h}}_{{\rm{U}},{K_{\rm{U}}}}}{\left( {{\bf{\tilde r}}} \right)}} \right] \in {\mathbb{C}^{{N_r} \times {K_{\rm{U}}}}}$, where ${{k_{\rm{U}}}} \in {{\cal K}_{\rm{U}}} \buildrel \Delta \over = \left\{ {k\left| {1 \le k \le {K_{\rm{U}}},k \in \mathbb{Z}} \right.} \right\}$. \\
	\indent Since the BS transmits and receives signals over the same time and frequency resource element, its transmitter introduce SI to its receiver. The SI channel, denoted as ${\bf{H}}_{{\rm{SI}}} \left( {{\bf{\tilde t}},{\bf{\tilde r}}} \right) \in {\mathbb{C}^{{N_r} \times {N_t}}}$, is determined by both the positions of transmit and receive PRAs. Conversely, DL users experience inter-user interference (IUI) from UL users, represented by ${\bf{H}}_{{\rm{IUI}}}\in {\mathbb{C}^{{K_{\rm{D}}} \times {K_{\rm{U}}}}}$. Each element ${h_{{\rm{IUI}},{k_{\rm{D}}},{k_{\rm{U}}}}}$ in the ${k_{\rm{D}}}$-th row and ${k_{\rm{U}}}$-th column of ${\bf{H}}_{{\rm{IUI}}}$ quantifies the interference from the ${k_{\rm{U}}}$-th UL user to the ${k_{\rm{D}}}$-th DL user. In this paper, we consider the scenario that the IUI power is comparable to that of the background noise, ensuring the FD communication remains feasible. \\
	\indent The received signal of the ${{k_{\rm{D}}}}$-th DL user is given by
	\begin{equation}\label{ykd}
		\begin{aligned}
			{y_{{\rm{D}},{k_{\rm{D}}}}}&={\bf{h}}_{{\rm{D}},{k_{\rm{D}}}}^{H}\left( {\tilde{\bf{t}}} \right){{\bf{w}}_{{t},{k_{\rm{D}}}}}{s_{{\rm{D}},{k_{\rm{D}}}}} + \sum\limits_{i = 1,i \ne {k_{\rm{D}}}}^{{K_{\rm{D}}}} {{\bf{h}}_{{\rm{D}},{k_{\rm{D}}}}^{H}\left( {\tilde{\bf{t}}} \right){{\bf{w}}_{{t},i}}{s_{{\rm{D}},i}}} \\
			&+\sum\limits_{{k_{\rm{U}}} = 1}^{{K_{\rm{U}}}} {\sqrt {{p_{{\rm{U}},{k_{\rm{U}}}}}} {h_{{\rm{IUI}},{k_{\rm{D}}},{k_{\rm{U}}}}}{s_{{\rm{U}},{k_{\rm{U}}}}}}  + {n_{{\rm{D}},{k_{\rm{D}}}}},
		\end{aligned}
	\end{equation}
	where ${{{\bf{W}}_{t}} = \left[ {{{\bf{w}}_{{t},1}}, \cdots ,{{\bf{w}}_{{t},{K_{\rm{D}}}}}} \right] \in {\mathbb{C}^{{N_t} \times {K_{\rm{D}}}}}}$ is the transmitter beamforming matrix with ${{{\bf{w}}_{{t}, {k_{\rm{D}}}}}} \in \mathbb{C}^{N_t \times 1}$ as the beamformer toward the ${{k_{\rm{D}}}}$-th DL user, the information vector to DL users ${{{\bf{s}}_{\rm{D}}} = \left[ {{s_{{\rm{D}},1}}, \cdots ,{s_{{\rm{D}},{K_{\rm{D}}}}}} \right] ^ T \in {\mathbb{C}^{{K_{\rm{D}}} \times 1}}}$ is with zero mean and identity matrix as covariance matrix. Similarly, ${{s_{{\rm{U}},{k_{\rm{U}}}}}}$ is the UL data with zero mean and unit variance of the ${{k_{\rm{U}}}}$-th UL user and ${{p_{{\rm{U}},{k_{\rm{U}}}}}}$ is the corresponding transmit power. ${n_{{\rm{D}},{k_{\rm{D}}}}}$ is the noise at the ${{k_{\rm{D}}}}$-th DL user following the circularly symmetric complex Gaussian (CSCG) distribution with zero mean and variance $\sigma_{{\rm{D}},{k_{\rm{D}}}}^2$, i.e., ${n_{{\rm{D}},{k_{\rm{D}}}}} \sim {\cal CN}\left( {0,\sigma _{{\rm{D}},{k_{\rm{D}}}}^2} \right)$. UL noise ${n_{{\rm{U}},{k_{\rm{U}}}}}$ has the same distribution. In (\ref{ykd}), the first term is the desired signal of the ${{k_{\rm{D}}}}$-th DL user and the other terms composed of MUI, IUI, and thermal noise are undesirable parts of the ${{k_{\rm{D}}}}$-th DL user. \\
	\indent The UL received signal at BS is represented as
	\begin{equation}
		{{\bf{y}}_{\rm{U}}} = {\bf{W}}_r^H \left( {{\bf{H}}_{\rm{U}}}\left( {{\bf{\tilde r}}} \right){{\bf{P}}_{\rm{U}}}{{\bf{s}}_{\rm{U}}} + {{\bf{H}}_{{\rm{SI}}}}\left( {{\bf{\tilde t}},{\bf{\tilde r}}} \right){{\bf{W}}_t}{{\bf{s}}_{\rm{D}}} + {{\bf{n}}_{\rm{U}}} \right), \notag
	\end{equation}
	where ${{\bf{y}}_{\rm{U}}} = \left[ {{y_{{\rm{U}},1}}, \cdots ,{y_{{\rm{U}},{K_{\rm{U}}}}}} \right] ^ T \in {\mathbb{C}^{{K_{\rm{U}}} \times 1}}$ is the signal collection of ${k_{\rm{D}}}$ UL users, ${{{\bf{W}}_{\rm{r}}} \in {\mathbb{C}^{{N_r} \times {K_{\rm{U}}}}}}$ is the stacked beamformer, ${{{\bf{P}}_{\rm{U}}} = {\rm{diag}}\left\{ {{p_{{\rm{U}},1}}, \cdots ,{p_{{\rm{U}},{K_{\rm{U}}}}}} \right\}}$ is the diagonal power matrix with the ${k_{\rm{U}}}$-th element as the transmit signal power of the ${{k_{\rm{U}}}}$-th UL user, and ${{{\bf{s}}_{\rm{U}}} = \left[ {{s_{{\rm{U}},1}}, \cdots ,{s_{{\rm{U}},{K_{\rm{U}}}}}} \right] ^ T \in {\mathbb{C}^{{K_{\rm{U}}} \times 1}}}$. Then, the received signal from the ${{k_{\rm{U}}}}$-th UL user is given by
	\begin{equation}\label{yku}
		\begin{aligned}
			{y_{{\rm{U}},{k_{\rm{U}}}}} &= \sqrt {{p_{{\rm{U}},{k_{\rm{U}}}}}} {\bf{w}}_{r,{k_{\rm{U}}}}^H{{\bf{h}}_{{\rm{U}},{k_{\rm{U}}}}}\left( {{\bf{\tilde r}}} \right){s_{{\rm{U}},{k_{\rm{U}}}}} \\
			&+ \sum\limits_{i = 1,i \ne {k_{\rm{U}}}}^{{K_{\rm{U}}}} {\sqrt {{p_{{\rm{U}},i}}} {\bf{w}}_{r,{k_{\rm{U}}}}^H{{\bf{h}}_{{\rm{U}},i}}\left( {{\bf{\tilde r}}} \right){s_{{\rm{U}},i}}} \\
			&+{\bf{w}}_{r,{k_{\rm{U}}}}^H{{\bf{H}}_{{\rm{SI}}}}\left( {{\bf{\tilde t}},{\bf{\tilde r}}} \right){{\bf{W}}_t}{{\bf{s}}_{\rm{D}}} + {\bf{w}}_{r,{k_{\rm{U}}}}^H{{\bf{n}}_{\rm{U}}}.
		\end{aligned}
	\end{equation}
	The first term of (\ref{yku}) is the desired signal from the ${{k_{\rm{U}}}}$-th UL user while the remaining terms represent the MUI, SI, and thermal noise.
	
	\subsection{Channel Model}
\begin{figure}[tb]
		\centering
		\includegraphics[scale=0.25]{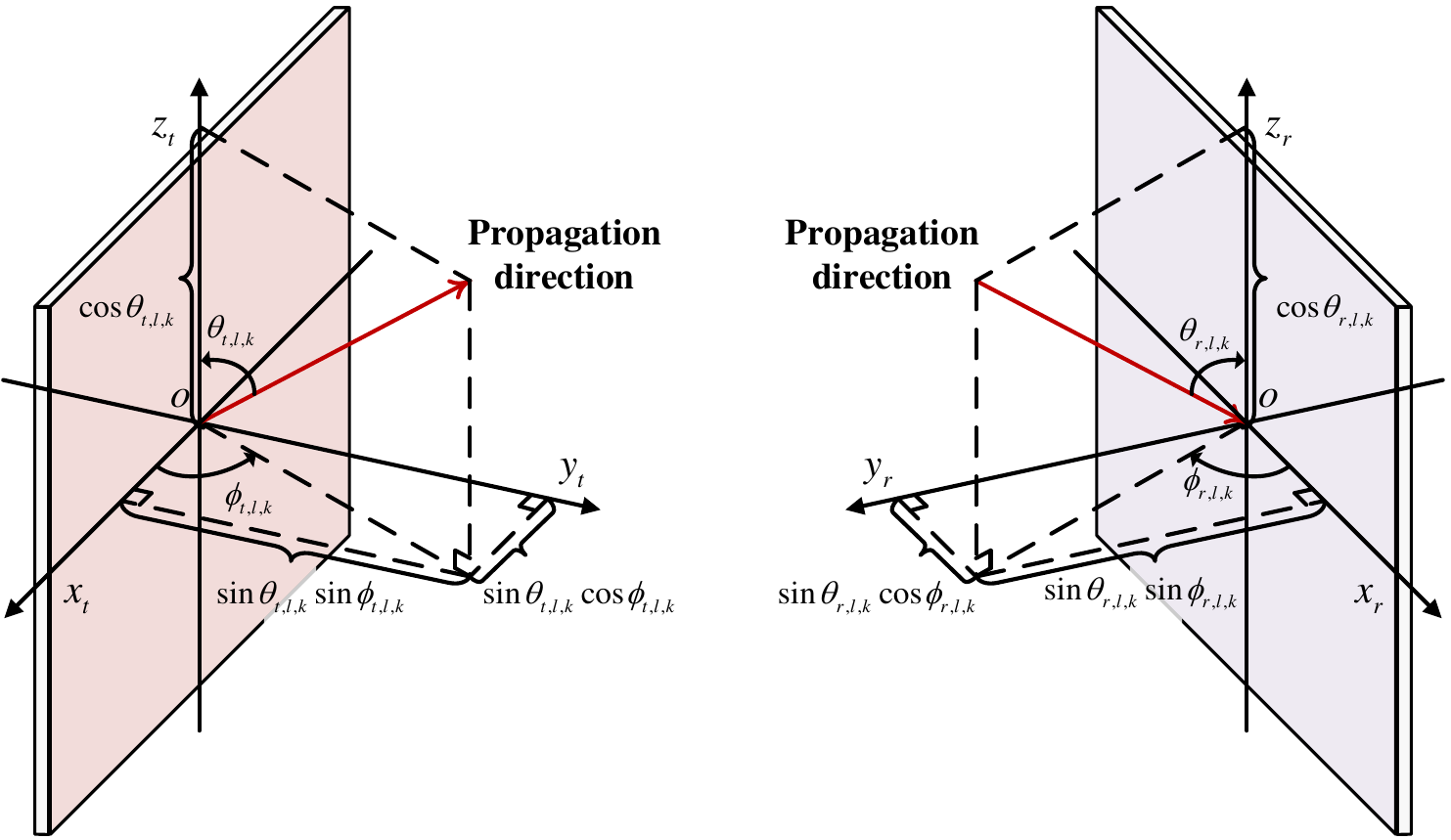}
		\caption{Spatial angles for transmit and	receive regions.}	
		\label{fig2}
		\vspace{-1cm}
	\end{figure}

	\indent Channel matrices are determined by the signal‐propagation environment and the spatial arrangement of the PRAs. \blue{In this paper, we consider low-mobility scenarios under far-field propagation conditions, which exhibit quasi-static and slow-fading channels with a long coherence time, such that the antenna positions are not required to adjust frequently. A finite-scattering channel model commonly adopted in relevant investigations \cite{MAS2, MAS-SFD-MU} is adopted. This assumption is widely used to represents scenarios like fixed wireless access, indoor communications, or backhaul/fronthaul links, where the propagation conditions are relatively stable.} We assume that the DL transmit channel from the BS to the ${{k_{\rm{D}}}}$-th DL user consists of ${L_{t,{k_{\rm{D}}}}}$ resolvable paths, while the UL receive channel from the ${{k_{\rm{U}}}}$-th UL user to the BS receiver has ${L_{r,{k_{\rm{U}}}}}$ resolvable paths. Furthermore, since we consider far-field scenarios in this paper, the size of transmit/receive reconfigurable region can be regarded as much smaller than the signal propagation distance. As a result, all PRAs at the BS transmitter/receiver experience equal path loss and share a common angle of departure (AoD)/angle of arrival (AoA) for each channel path. We define the elevation and azimuth AoDs of the ${l_t}$-th in the DL channel to the ${{k_{\rm{D}}}}$-th user as ${\theta _{t,{l_t},{k_{\rm{D}}}}}, {\phi _{t,{l_t},{k_{\rm{D}}}}} \in \left[ {0,\pi } \right]$. Similarly, we can get the set of number of path and corresponding AoD or AoA for all DL and UL channels as $\left\{ {{L_{t,{k_{\rm{D}}}}},\left\{ {{\theta _{t,{l_t},{k_{\rm{D}}}}}} \right\}_{{l_t} = 1}^{{L_{t,{k_{\rm{D}}}}}},\left\{ {{\phi _{t,{l_t},{k_{\rm{D}}}}}} \right\}_{{l_t} = 1}^{{L_{t,{k_{\rm{D}}}}}}} \right\}_{{k_{\rm{D}}} = 1}^{{K_{\rm{D}}}}$ and $\left\{ {{L_{r,{k_{\rm{U}}}}},\left\{ {{\theta _{r,{l_r},{k_{\rm{U}}}}}} \right\}_{{l_r} = 1}^{{L_{r,{k_{\rm{U}}}}}},\left\{ {{\phi _{r,{l_r},{k_{\rm{U}}}}}} \right\}_{{l_r} = 1}^{{L_{r,{k_{\rm{U}}}}}}} \right\}_{{k_{\rm{U}}} = 1}^{{K_{\rm{U}}}}$. The spatial angles are illustrated in Fig.\ref{fig2} based on these definition. From the illustration, the normalized wave vector of the $l_t$-th transmit path in the ${{k_{\rm{D}}}}$-th DL channel is calculated as ${{\bf{n}}_{t,{l_t},{{k_{\rm{D}}}}}} = {\left[ {\sin {\theta _{t,{l_t},{{k_{\rm{D}}}}}}\cos {\phi _{t,{l_t},{{k_{\rm{D}}}}}},\cos {\theta _{t,{l_t},{{k_{\rm{D}}}}}}} \right]^T}$. \\
	\indent Although all PRAs are assume to share the same amplitude of channel coefficient, they experience signal with different phase for every path. Taking the origin for reference, the signal propagation distance difference of the $n_t$-th transmit PRA in the $l_{t,{k_{\rm{D}}}}$-th path is given by 
	\begin{align}
		&{\rho _{t,{l_t},{k_{\rm{D}}}}} \left( {{{\bf{t}}_{{n_t}}}} \right) = {\bf{n}}_{t,{l_t},{k_{\rm{D}}}}^T{{\bf{t}}_{{n_t}}}  \notag \\
		&= {x_{{t,n_t}}}\sin {\theta _{t,{l_t},{k_{\rm{D}}}}}\cos {\phi _{t,{l_t},{k_{\rm{D}}}}} + {y_{{t,n_t}}}\cos {\theta _{t,{l_t},{k_{\rm{D}}}}}. \notag
	\end{align}
	The phase difference is then $ 2\pi {{{\rho _{t,{l_t},{k_{\rm{D}}}}}\left( {{{\bf{t}}_{{n_t}}}} \right)} / \lambda }$, where $\lambda$ is the carrier wavelength. The field response vector of the $n_t$-th transmit PRA belonging to ${\mathbb{C}^{{L_t} \times 1}}$ can be represented as 
	\begin{equation}
		{{\bf{g}}_{k_{\rm{D}}}}\left( {{{\bf{t}}_{{n_t}}}} \right) \buildrel \Delta \over = {\left[ {{e^{j\frac{{2\pi }}{\lambda }{\rho _{t,1,{k_{\rm{D}}}}}\left( {{{\bf{t}}_{{n_t}}}} \right)}}, \cdots ,{e^{j\frac{{2\pi }}{\lambda }{\rho _{t,{L_t},{k_{\rm{D}}}}}\left( {{{\bf{t}}_{{n_t}}}} \right)}}} \right]^T}.\notag
	\end{equation}
	Collecting the field response vectors of all transmit PRAs, the field response matrix of transmitter is given by
	\begin{equation}
		{{\bf{G}}_{k_{\rm{D}}}}\left( {{\bf{\tilde t}}} \right) \buildrel \Delta \over = \left[ {{{\bf{g}}_{k_{\rm{D}}}}\left( {{{\bf{t}}_1}} \right), \cdots ,{{\bf{g}}_{k_{\rm{D}}}}\left( {{{\bf{t}}_{{N_t}}}} \right)} \right] \in {\mathbb{C}^{{L_t} \times {N_t}}}.\notag
	\end{equation}
	The field response vector of the $n_r$-th receive PRA and the field response matrix of receiver  for the ${{k_{\rm{U}}}}$-th UL user belonging to ${\mathbb{C}^{{L_r} \times 1}}$ and ${\mathbb{C}^{{L_r} \times {N_r}}}$ are respectively given by
	\begin{equation}
		{{\bf{f}}_{k_{\rm{U}}}}\left( {{{\bf{r}}_{{n_r}}}} \right) \buildrel \Delta \over = {\left[ {{e^{j\frac{{2\pi }}{\lambda }{\rho _{r,1,{k_{\rm{U}}}}}\left( {{{\bf{r}}_{{n_r}}}} \right)}}, \cdots ,{e^{j\frac{{2\pi }}{\lambda }{\rho _{r,{L_r},{k_{\rm{U}}}}}\left( {{{\bf{r}}_{{n_r}}}} \right)}}} \right]^T}; \notag
	\end{equation}
	and
	\begin{equation}
		{{\bf{F}}_{k_{\rm{U}}}}\left( {{\bf{\tilde r}}} \right) \buildrel \Delta \over = \left[ {{\bf{f}}_{k_{\rm{U}}} \left( {{{\bf{r}}_1}} \right), \cdots ,{\bf{f}}_{k_{\rm{U}}} \left( {{{\bf{r}}_{{N_r}}}} \right)} \right].\notag
	\end{equation}
	In a similar way, we define the field response vectors and matrices of SI channel ${{\bf{g}}_{{\rm{SI}}}}\left( {{{\bf{t}}_{{n_t}}}} \right) \in {\mathbb{C}^{{L_{t,{\rm{SI}}}} \times 1}}$, ${{\bf{f}}_{{\rm{SI}}}}\left( {{{\bf{r}}_{{n_r}}}} \right) \in {\mathbb{C}^{{L_{r,{\rm{SI}}}} \times 1}}$, ${{\bf{G}}_{{\rm{SI}}}}\left( {{\bf{\tilde t}}} \right) \in {\mathbb{C}^{{L_{t,{\rm{SI}}}} \times {N_t}}}$, and ${{\bf{F}}_{{\rm{SI}}}}\left( {{\bf{\tilde r}}} \right) \in {C^{{L_{r,{\rm{SI}}}} \times {N_r}}}$.
	To this end, channels in (\ref{ykd}) and (\ref{yku}) can be represented as
	\begin{align}
		{{\bf{h}}_{{\rm{D}},{k_{\rm{D}}}}}\left( {{\bf{\tilde t}}} \right) &= {\bf{G}}_{{k_{\rm{D}}}}^H\left( {{\bf{\tilde t}}} \right){\Sigma _{{\rm{D}},{k_{\rm{D}}}}}{{\bf{1}}_{{L_{t,{k_{\rm{D}}}}}}},\notag \\
		{{\bf{h}}_{{\rm{U}},{k_{\rm{U}}}}}\left( {{\bf{\tilde r}}} \right) &= {\bf{F}}_{{k_{\rm{U}}}}^H\left( {{\bf{\tilde r}}} \right){\Sigma _{{\rm{U}},{k_{\rm{U}}}}}{{\bf{1}}_{{L_{r,{k_{\rm{U}}}}}}},\notag \\
		{{\bf{H}}_{{\rm{SI}}}}\left( {{\bf{\tilde t}},{\bf{\tilde r}}} \right) &= {\bf{F}}_{{\rm{SI}}}^H\left( {{\bf{\tilde r}}} \right){\Sigma _{{\rm{SI}}}}{{\bf{G}}_{{\rm{SI}}}}\left( {{\bf{\tilde t}}} \right),\notag
	\end{align}
	where ${\Sigma _{{\rm{D}},{k_{\rm{D}}}}} \in \mathbb{C}^{L_{t,{k_{\rm{D}}}} \times L_{t,{k_{\rm{D}}}}}$, $\Sigma _{{\rm{U}},{k_{\rm{U}}}} \in \mathbb{C}^{L_{r,{k_{\rm{U}}}} \times L_{r,{k_{\rm{U}}}}}$, $\Sigma_{{\rm{SI}}} \in \mathbb{C}^{{L_{r,{{\rm{SI}}}}} \times {L_{t,{{\rm{SI}}}}}}$ are the path response matrices (PRMs), and ${\bf{1}}_m$ is an all-one vector with the length of $m$. In this paper, we consider a quasi-static block-fading channel and focus on a single fading block. That is, within any given location in ${\cal C}_t/{\cal C}_r$, the multi-path channel components remain fixed, and the PRMs are considered constant.
	\subsection{Problem Formulation}
	From Eq.(\ref{ykd}) and (\ref{yku}), the DL and UL signal-to-interference-plus-noise ratios (SINR) are given by (\ref{sinr1}) and (\ref{sinr2}), as shown in the next page. Correspondingly, the achievable rates of the ${{k_{\rm{D}}}}$-th DL and the ${{k_{\rm{U}}}}$-th UL user can be estimated as ${R_{{\rm{D}},{k_{\rm{D}}}}} = {\log _2}\left( {1 + {\gamma _{{\rm{D}},{k_{\rm{D}}}}}} \right)$ and ${R_{{\rm{U}},{k_{\rm{U}}}}} = {\log _2}\left( {1 + {\gamma _{{\rm{U}},{k_{\rm{U}}}}}} \right)$, respectively. Our objective is to maximize the weighted sum of the achievable rates for all DL and UL users incorporating a serving priority set $\left\{ {{a_i}} \right\}_{i = 1}^{{K_{\rm{D}}} + {K_{\rm{U}}}}$, where ${\sum\limits_{i = 1}^{{K_{\rm{D}}} + {K_{\rm{U}}}} {{a_i} = 1} }$. This is achieved through a joint optimization over the PRA positions, DL and UL beamformers, and power allocation. The corresponding optimization problem is formulated as follows
	\begin{align}
		\left( {\rm{P1}}\right)  \  &\mathop {\max }\limits_{{\bf{\tilde t}},{\bf{\tilde r}},{{\bf{W}}_t},{{\bf{W}}_r},{{\bf{P}}_{\rm{U}}}} \  \sum\limits_{{k_{\rm{D}}} = 1}^{{K_{\rm{D}}}} {{a_{{k_{\rm{D}}}}}{R_{{\rm{D}},{k_{\rm{D}}}}}}  + \sum\limits_{{k_{\rm{U}}} = 1}^{{K_{\rm{U}}}} {{a_{{K_{\rm{D}}} + {k_{\rm{U}}}}}{R_{{\rm{U}},{k_{\rm{U}}}}}} \label{1} \\
		\text{s.t.} \ \  &{{{\bf{t}}_{{n_t}}} \in {{\cal C}_t},{{\bf{r}}_{{n_r}}} \in {{\cal C}_r},{n_t} \in {{\cal N}_t},{n_r} \in {{\cal N}_r}}, \tag{\ref{1}{a}} \label{1a}\\
		&{{{\left\| {{{\bf{t}}_{{n_1}}} - {{\bf{t}}_{{n_2}}}} \right\|}_2} \ge {D_{\min }},{n_1, n_2} \in {{\cal N}_t},{n_1} \ne {n_2}}, \tag{\ref{1}{b}} \label{1b}\\
		&{{{\left\| {{{\bf{r}}_{{n_1}}} - {{\bf{r}}_{{n_2}}}} \right\|}_2} \ge {D_{\min }},{n_1, n_2} \in {{\cal N}_r},{n_1} \ne {n_2}}, \tag{\ref{1}{c}} \label{1c}\\
		&{\rm{Tr}}\left( {{{\bf{W}}_t}{\bf{W}}_t^H} \right) \le {p_{{\rm{D,max}}}}, \tag{\ref{1}{d}} \label{1d} \\
		&{\left\| {{\bf{w}}_{r,{k_{\rm{U}}}}} \right\|_2^2 = 1}, {k_{\rm{U}}} \in {{\cal K}_{\rm{U}}}, \tag{\ref{1}{e}} \label{1e}\\
		&0 \le {p_{{\rm{U}},{k_{\rm{U}}}}} \le {p_{{\rm{U,max}}}}, {k_{\rm{U}}} \in {{\cal K}_{\rm{U}}}. \tag{\ref{1}{f}} \label{1f}
	\end{align}
	In (P1), constraints (\ref{1a})-(\ref{1c}) regulate the reconfigurable region and enforce the minimum inter-antenna distance. Meanwhile, constraints (\ref{1d})-(\ref{1f}) impose limitations on the power budget. \blue{Note that we assume the SI is substantially mitigated using techniques mentioned in \cite{FD1} and \cite{FD2} and the DL and UL users are not closely located. Hence, the power of SI and IUI satisfies the operational requirement.}
	\section{Proposed Solution}
	The objective function of (P1) follows a sum-log-ratio form with highly coupled designed variables. Additionally, the constraints on antenna placement create an irregular and non-convex feasible region for antenna positioning. While the power constraints remain convex, the requirement to regulate the receive beamformer on a spherical surface introduces non-convexity as well. Consequently, (P1) is intractable and presents a highly non-convex optimization challenge. \blue{For such a highly non-convex and coupled problem, AO is an effective framework to solve this problem.} Hence, in this section, we propose an AO algorithm to solve the formulated (P1). By decomposing the original complex problem into several more manageable sub-problems, we can iteratively optimize each variable while gradually addressing the non-convex objective function and constraints. \\
	\indent We first eliminate the constraint (\ref{1e}) to simplify the optimization problem and reduce the complexity of the feasible solution space. \\
	\indent \textit{Proposition 1: Scaling ${{{\bf{w}}_{r,{k_{\rm{U}}}}}}$ holds the optimality of (P1).} \\
	\indent Proof: It is noticed that only ${{R_{{\rm{U}},{k_{\rm{U}}}}}}$ term in objective function involves ${{\bf{w}}_{r,{k_{\rm{U}}}}}$. Consequently, we focus on $R_{\rm{U}}$ and assume that the optimal solution for (P1) without (\ref{1e}) is ${{{\bf{\tilde t}}}^{\star}}$, ${{{\bf{\tilde r}}}^{\star}}$, ${{\bf{P}}_{\rm{U}}^\star}$ ${\bf{W}}_t^{\star}$, and ${\bf{W}}_r^{\star}$ (not necessarily on the unit manifold now) and the optimal value for UL part is $R_{\rm{U}}^{\star}$, $R_{{\rm{U}},{k_{\rm{U}}}}^{\star}$ for every UL user respectively. For any $t \in \mathbb{R}$, plugging $t{\bf{w}}_{r,{k_{\rm{U}}}}^{\star}$ into the expression of ${{R_{{\rm{U}},{k_{\rm{U}}}}}}$, it is easily obtained that ${R_{{\rm{U}},{k_{\rm{U}}}}}\left( {t{\bf{w}}_{r,{k_{\rm{U}}}}^{\star}} \right) = R_{{\rm{U}},{k_{\rm{U}}}}^{\star}$, i.e., ${R_{{\rm{U}},{k_{\rm{U}}}}}$ remains its value with scaled ${{\bf{w}}_{r,{k_{\rm{U}}}}}$. Each UL user's achievable rate is invariant with respect to (w.r.t.) the scaling of beamformer and thus, the weighted-sum rate also keeps its value. In other words, scaling ${{\bf{w}}_{r,{k_{\rm{U}}}}}$ holds the optimality. $\hfill\blacksquare$ \\
	\indent \textit{Proposition 1} allows us to neglect (\ref{1e}) when solving (P1) and the optimal ${{\bf{w}}_{r,{k_{\rm{U}}}}}$ can be acquired by a column-wise normalization on ${\bf{W}}_r$. \\
	\indent The sum-log-ratio form in the objective function of (P1) falls within the general domain of fractional programming (FP). To the best of our knowledge, no existing methods can directly solve this type of multiple-ratio FP within a feasible time. Even finding stationary-point solutions is challenging, apart from general-purpose techniques such as SCA. The method in \cite{FP1, FP2} addresses this kind of problem from a new perspective by introducing auxiliary variables.
	
	\vspace{-0.5cm}
	\subsection{Lagrangian dual transform}
	\begin{figure*}[b]
		\centering
		\hrulefill
		\begin{subequations}
			\begin{align}
				&{\gamma _{{\rm{D}},{k_{\rm{D}}}}} = {{\left| {{\bf{h}}_{{\rm{D}},{k_{\rm{D}}}}^{H}\left( {{\bf{\tilde t}}} \right){{\bf{w}}_{t,{k_{\rm{D}}}}}} \right|}^2} / \left( {\sum\limits_{i = 1,i \ne {k_{\rm{D}}}}^{{K_{\rm{D}}}} {{{\left| {{\bf{h}}_{{\rm{D}},{k_{\rm{D}}}}^{H}\left( {{\bf{\tilde t}}} \right){{\bf{w}}_{t,i}}} \right|}^2}}  + \sum\limits_{{k_{\rm{U}}} = 1}^{{K_{\rm{U}}}} {{{\left| {{h_{{\rm{IUI}},{k_{\rm{D}}},{k_{\rm{U}}}}}} \right|}^2}{p_{{\rm{U}},{k_{\rm{U}}}}}}  + \sigma _{{\rm{D}},{k_{\rm{D}}}}^2}\right), \label{sinr1} \\
				&{\gamma _{{\rm{U}},{k_{\rm{U}}}}} = {{\left| {{\bf{w}}_{r,{k_{\rm{U}}}}^H{{\bf{h}}_{{\rm{U}},{k_{\rm{U}}}}}\left( {{\bf{\tilde r}}} \right)} \right|}^2}{p_{{\rm{U}},{k_{\rm{U}}}}} / \left( {\sum\limits_{i = 1,i \ne {k_{\rm{U}}}}^{{K_{\rm{U}}}} {{{\left| {{\bf{w}}_{r,{k_{\rm{U}}}}^H{{\bf{h}}_{{\rm{U}},i}}\left( {{\bf{\tilde r}}} \right)} \right|}^2}{p_{{\rm{U}},i}}}  + \left| {{\bf{w}}_{r,{k_{\rm{U}}}}^H{{\bf{H}}_{{\rm{SI}}}}\left( {{\bf{\tilde t}},{\bf{\tilde r}}} \right){{\bf{W}}_t}} \right|^2 + \left\| {{{\bf{w}}_{r,{k_{\rm{U}}}}}} \right\|_2^2\sigma _{{\rm{U}}}^2}\right). \label{sinr2}
			\end{align}
		\end{subequations}
		\begin{equation} \label{gamma}
			\gamma _i^{\star} = \left\{ \begin{array}{l}
				{{\left| {{\bf{h}}_{{\rm{D}},i}^{H}\left( {{\bf{\tilde t}}} \right){{\bf{w}}_{t,i}}} \right|}^2} / \left( {\sum\limits_{j = 1,j \ne i}^{{K_{\rm{D}}}} {{{\left| {{\bf{h}}_{{\rm{D}},i}^{H}\left( {{\bf{\tilde t}}} \right){{\bf{w}}_{t,j}}} \right|}^2}}  + \sum\limits_{{k_{\rm{U}}} = 1}^{{K_{\rm{U}}}} {{{\left| {{h_{{\rm{IUI}},i,{k_{\rm{U}}}}}} \right|}^2}{p_{{\rm{U}},{k_{\rm{U}}}}}}  + \sigma _{{\rm{D}},i}^2}\right)  \qquad \qquad \qquad \quad \ ,1 \le i \le {K_{\rm{D}}}\\
				\frac{{{{\left| {{\bf{w}}_{r,i - {K_{\rm{D}}}}^H{{\bf{h}}_{{\rm{U}},i - {K_{\rm{D}}}}}\left( {{\bf{\tilde r}}} \right)} \right|}^2}{p_{{\rm{U}},i - {K_{\rm{D}}}}}}}{{\sum\limits_{j = 1,j \ne i - {K_{\rm{D}}}}^{{K_{\rm{U}}}} {{p_{{\rm{U}},j}}{{\left| {{\bf{w}}_{r,i - {K_{\rm{D}}}}^H{{\bf{h}}_{{\rm{U}},j}}\left( {{\bf{\tilde r}}} \right)} \right|}^2}}  + {{{\left\| {{\bf{w}}_{r,i - {K_{\rm{D}}}}^H{{\bf{H}}_{{\rm{SI}}}}\left( {{\bf{\tilde t}},{\bf{\tilde r}}} \right){{\bf{W}}_t}} \right\|}^2}} + \left\| {{{\bf{w}}_{r,i - {K_{\rm{D}}}}}} \right\|_2^2\sigma _{\rm{U}}^2}}\qquad \qquad \quad ,{K_{\rm{D}}} + 1 \le i \le {K_{\rm{D}}} + {K_{\rm{U}}}
			\end{array} \right.
		\end{equation}
	\end{figure*}
	\blue{The Lagrangian dual transform introduces a non-negative real auxiliary variable $\bm{\gamma} \in \mathbb{R}_ + ^{\left( {{K_{\rm{D}}} + {K_{\rm{U}}}} \right) \times 1}$ to explicitly represent the effective SINR, which decouples the signal and interference terms inside the logarithm and enables linearization of the objective.} The new problem is then given by
	\begin{align}
		\left( {\rm{P2}}\right)  \  \mathop {\max }\limits_{{\bf{\tilde t}},{\bf{\tilde r}},{{\bf{W}}_t},{{\bf{W}}_r},{{\bf{P}}_{\rm{U}}},{\bm{\gamma}}} &\sum\limits_{i = 1}^{{K_{\rm{D}}} + {K_{\rm{U}}}} {{a_i}\left( {\log \left( {1 + {\gamma _i}} \right) - {\gamma _i}} \right)}  + {S_1} + {S_2} \notag  \\
		\text{s.t.} \qquad &{\bm{\gamma}} \succeq 0, (\ref{1a}) - (\ref{1f}), \notag
	\end{align}
	where
	\begin{equation}
		{S_1} = \sum\limits_{{k_{\rm{D}}} = 1}^{{K_{\rm{D}}}} {{a_{{k_{\rm{D}}}}}\left( {1 + {\gamma _{{k_{\rm{D}}}}}} \right)\frac{{{{\left| {{\bf{h}}_{{\rm{D}},{k_{\rm{D}}}}^{H}\left( {{\bf{\tilde t}}} \right){{\bf{w}}_{t,{k_{\rm{D}}}}}} \right|}^2}}}{{{{s_{1,{k_{\rm{D}}}}}}}}},\notag
	\end{equation}
	\begin{align}
		{{s_{1,{k_{\rm{D}}}}}} &= \sum\limits_{i = 1}^{{K_{\rm{D}}}} {{\left| {{\bf{h}}_{{\rm{D}},{k_{\rm{D}}}}^{H}\left( {{\bf{\tilde t}}} \right){{\bf{w}}_{t,i}}} \right|}^2} + \sum\limits_{{k_{\rm{U}}} = 1}^{{K_{\rm{U}}}} {{{\left| {{h_{{\rm{IUI}},{k_{\rm{D}}},{k_{\rm{U}}}}}} \right|}^2}{p_{{\rm{U}},{k_{\rm{U}}}}}} \notag \\
		&+ \sigma _{{\rm{D}},{k_{\rm{D}}}}^2 \notag
	\end{align}
	\begin{equation}
		{S_2} = \sum\limits_{{k_{\rm{U}}} = 1}^{{K_{\rm{U}}}} {{a_{{K_{\rm{D}}} + {k_{\rm{U}}}}}{p_{{\rm{U}},{k_{\rm{U}}}}}\left( {1 + {\gamma _{{K_{\rm{D}}} + {k_{\rm{U}}}}}} \right)\frac{{{{\left| {{\bf{w}}_{r,{k_{\rm{U}}}}^H{{\bf{h}}_{{\rm{U}},{k_{\rm{U}}}}}\left( {{\bf{\tilde r}}} \right)} \right|}^2}}}{{{s_{2,{k_{\rm{U}}}}}}}},\notag
	\end{equation}
	and
	\begin{align}
		{s_{2,{k_{\rm{U}}}}} &= \sum\limits_{i = 1}^{{K_{\rm{U}}}} {{{\left| {{\bf{w}}_{r,{k_{\rm{U}}}}^H{{\bf{h}}_{{\rm{U}},i}}\left( {{\bf{\tilde r}}} \right)} \right|}^2}{p_{{\rm{U}},i}}}  + {{{\left\| {{\bf{w}}_{r,{k_{\rm{U}}}}^H{{\bf{H}}_{{\rm{SI}}}}\left( {{\bf{\tilde t}},{\bf{\tilde r}}} \right){{\bf{W}}_t}} \right\|}^2}} \notag \\
		&+ \left\| {{{\bf{w}}_{r,{k_{\rm{U}}}}}} \right\|_2^2\sigma _{\rm{U}}^2.\notag
	\end{align}
	The equivalence of (P2) to (P1) can be referred to the proof in \cite{FP2}. Since (P2) is convex over $\bm{\gamma}$, we can obtain the optimal $\bm{\gamma}$ in a closed-form expression when ${\bf{\tilde t}}$, ${\bf{\tilde r}}$, ${{\bf{W}}_t}$, ${{\bf{W}}_r}$, and ${{\bf{P}}_{\rm{U}}}$ are fixed. The optimal ${\bm{\gamma}}$ is determined as (\ref{gamma}) below.
	
	\vspace{-0.3cm}
	\subsection{Quadratic transform}
	\blue{In the quadratic transform, the second auxiliary variable ${\bf{y}} \in \mathbb{C} ^{\left( {{K_{\rm{D}}} + {K_{\rm{U}}}} \right) \times 1}$ is introduced as a receive filter, reformulating the fractional term into a quadratic expression that is convex in each variable. By introducing ${\bf{y}}$, the quadratic transform decouples the mutual coupling between optimization variables in ratio expressions, thereby rendering the problem more tractable.} The newly transformed problem is given by
	\begin{align}
		\left( {\rm{P3}}\right)  \  &\mathop {\max }\limits_{\scriptstyle{\bf{\tilde t}},{\bf{\tilde r}},{{\bf{W}}_t},{{\bf{W}}_r},\hfill\atop
			\ \ \scriptstyle{{\bf{P}}_{\rm{U}}},{\bf{\gamma }},{\bf{y}}\hfill} \sum\limits_{i = 1}^{{K_{\rm{D}}} + {K_{\rm{U}}}} {{a_i}\left( {\log \left( {1 + {\gamma _i}} \right) - {\gamma _i}} \right)}  + {T_1} + {T_2} \notag  \\
		&\text{s.t.} \quad {\bm{\gamma}} \succeq 0, (\ref{1a}) - (\ref{1f}),  \notag
	\end{align} 
	where
	\begin{align}
		{T_1} &= \sum\limits_{{k_{\rm{D}}} = 1}^{{K_{\rm{D}}}} 2\sqrt {{a_{{k_{\rm{D}}}}}\left( {1 + {\gamma _{{k_{\rm{D}}}}}} \right)} {\mathop{\rm Re}\nolimits} \left\{ {{y_{{k_{\rm{D}}}}^*}{\bf{h}}_{{\rm{D}},{k_{\rm{D}}}}^{H}\left( {{\bf{\tilde t}}} \right){{\bf{w}}_{t,{k_{\rm{D}}}}}} \right\} \notag \\ 
		&- {{\left| {{y_{{k_{\rm{D}}}}}} \right|}^2}{s_{1,{k_{\rm{D}}}}}, \notag
	\end{align}
	\begin{align}
		{T_2} &= \sum\limits_{{k_{\rm{U}}} = 1}^{{K_{\rm{U}}}}  2\sqrt {{a_{{K_{\rm{D}}} + {k_{\rm{U}}}}}{p_{{\rm{U}},{k_{\rm{U}}}}}\left( {1 + {\gamma _{{K_{\rm{D}}} + {k_{\rm{U}}}}}} \right)} \cdot \notag \\
		&{\mathop{\rm Re}\nolimits} \left\{ {{y_{{K_{\rm{D}}} + {k_{\rm{U}}}}^*}{\bf{w}}_{r,{k_{\rm{U}}}}^H{{\bf{h}}_{{\rm{U}},{k_{\rm{U}}}}}\left( {{\bf{\tilde r}}} \right)} \right\} - {{\left| {{y_{{K_{\rm{D}}} + {k_{\rm{U}}}}}} \right|}^2}{s_{2,{k_{\rm{U}}}}}. \notag
	\end{align}
	(P3) is equivalent to (P2) \cite{FP1} with optimal $\bf{y}^{\star}$:
	\begin{equation} \label{y}
		{y_i^{\star}} = \left\{ \begin{array}{l}
			\frac{{\sqrt {{a_i}\left( {1 + {\gamma _i}} \right)} {\bf{h}}_{{\rm{D}},i}^H\left( {{\bf{\tilde t}}} \right){{\bf{w}}_{t,i}}}}{{{s_{1,i}}}} \qquad \qquad \quad \quad \ ,1 \le i \le {K_{\rm{D}}},\\
			\frac{{\sqrt {{a_i}{p_{{\rm{U}},i - {K_{\rm{D}}}}}\left( {1 + {\gamma _i}} \right)} {\bf{h}}_{{\rm{U}},i - {K_{\rm{D}}}}^H\left( {{\bf{\tilde t}}} \right){{\bf{w}}_{r,i - {K_{\rm{D}}}}}}}{{{s_{2,i}}}}, \text{otherwise},
		\end{array} \right.
	\end{equation}
	which is acquired from the first-order optimality condition. \blue{The above two introduced transformations remove the difficulty of directly optimizing non-convex ratios, provide closed-form updates in each iteration, and allow the overall problem to be solved efficiently via alternating optimization.}
	\vspace{-0.15cm}
	\subsection{Sub-problem 1: Optimizing ${\bf{W}}_t$}
	Given ${{\bf{W}}_r}$, ${{\bf{P}}_{\rm{U}}}$, ${\bf{\tilde t}}$, ${\bf{\tilde r}}$, ${\bf{y}}$, and $\boldsymbol{\gamma}$, (P3) degrades to 
	\begin{align}
		\left( {\rm{P4}}\right)  \quad \mathop {\max }\limits_{{{\bf{W}}_t}} \quad &2{\mathop{\rm Re}\nolimits} \left\{ {{\rm{Tr}}\left( {{\bf{\bar H}}_{\rm{D}}^H{{\bf{W}}_t}} \right)} \right\} - {\rm{Tr}}\left( {{\bf{W}}_t^H{{\bf{H}}_t}{{\bf{W}}_t}} \right) \label{sp1} \\
		\text{s.t.} \quad &(\ref{1d}), \notag
	\end{align}
	\blue{where}
	$${{{\bf{\bar H}}}_{\rm{D}}} = {{\bf{H}}_{\rm{D}}}{{\bf{Y}}_{\rm{D}}}{{\bf{A}}_{\rm{D}}},$$ $${{\bf{H}}_t} = {{\bf{H}}_{\rm{D}}}{{\bf{Y}}_{\rm{D}}}{\bf{Y}}_{\rm{D}}^H{\bf{H}}_{\rm{D}}^H + {\bf{H}}_{{\rm{SI}}}^H{{\bf{W}}_r}{\bf{Y}}_{\rm{U}}^H{{\bf{Y}}_{\rm{U}}}{\bf{W}}_r^H{{\bf{H}}_{{\rm{SI}}}}.$$
	In detail, ${{\bf{Y}}_{\rm{D}}} = {\rm{diag}}\left( {\left\{ {{y_i}} \right\}_{i = 1}^{{K_{\rm{D}}}}} \right)$, ${{\bf{Y}}_{\rm{U}}} = {\rm{diag}}\left( {\left\{ {{y_i}} \right\}_{i = {K_{\rm{D}}} + 1}^{{K_{\rm{D}}} + {K_{\rm{U}}}}} \right)$, and ${\bf{A}_{\rm{D}}} = {\rm{diag}}\left( {\left\{ {\sqrt {{a_i}\left( {1 + {\gamma _i}} \right)} } \right\}_{i = 1}^{{K_{\rm{D}}}}} \right)$. Since the DL beamformers ${{{\bf{W}}_t}} $ are not restricted to unit power and their sum-power is constrained by (\ref{1d}), (P4) is essentially a combination of beamforming and power allocation. From an optimization theory perspective, (P4) is a quadratically constrained quadratic optimization problem with a positive semi-definite matrix ${{\bf{H}}_t}$. Given this structure, the problem (\ref{sp1}) is convex and can be directly solved using various optimization tools, e.g. CVX toolbox in Matlab. \\
	\indent However, to reduce computational complexity, we propose to further break (P4) using a Lagrange multiplier method. The Lagrangian function of (P4) is expressed as 
	\begin{align}
		L\left( {{{\bf{W}}_t},\lambda } \right) &= {\rm{Tr}}\left( {{\bf{W}}_t^H{{\bf{H}}_t}{{\bf{W}}_t}} \right) - 2{\mathop{\rm Re}\nolimits} \left\{ {{\rm{Tr}}\left( {{\bf{\bar H}}_{\rm{D}}^H{{\bf{W}}_t}} \right)} \right\} \notag \\
		&+ \lambda \left( {{\rm{Tr}}\left( {{\bf{W}}_t^H{{\bf{W}}_t}} \right) - {p_{{\rm{D,max}}}}} \right),\notag
	\end{align}
	which yields the Karush-Kuhn-Tucker (KKT) condition
	\begin{equation} \label{KKT}
		\left\{ \begin{array}{l}
			{\rm{Tr}}\left( {{{\left( {{\bf{W}}_t^*} \right)}^H}{\bf{W}}_t^*} \right) \le {p_{{\rm{D,max}}}}, {\mu ^*} \ge 0,\\
			{\mu ^*}\left( {{\rm{Tr}}\left( {{{\left( {{\bf{W}}_t^*} \right)}^H}{\bf{W}}_t^*} \right) - {p_{{\rm{D,max}}}}} \right) = 0,\\
			{\bf{W}}_t^* = {\left( {{{\bf{H}}_t} + {\mu ^*}{\bf{I}}} \right)^{ - 1}}{{{\bf{\bar H}}}_{\rm{D}}},
		\end{array} \right.
	\end{equation}
	where ${{\bf{W}}_t^{\star}}$ and $\mu^{\star}$ are the optimal solutions of (P4) and its dual problem, respectively, because of the convexity of (P4). If ${\rm{Tr}}\left( {{\bf{\bar H}}_{\rm{D}}^H{\bf{H}}_t^{ - 2}{{{\bf{\bar H}}}_{\rm{D}}}} \right) = {p_{{\rm{D,max}}}}$, then $\mu^\star=0$ and ${\bf{W}}_t^\star = {\bf{H}}_t^{ - 1}{{{\bf{\bar H}}}_{\rm{D}}}$. Otherwise, expressing ${{\bf{W}}_t^{\star}}$ with $\mu^\star$, we have 
	\begin{align}
		&{\rm{Tr}}\left( {{{\left( {{\bf{W}}_t^\star\left( {{\mu ^\star}} \right)} \right)}^H}{\bf{W}}_t^\star\left( {{\mu ^\star}} \right)} \right) \notag \\
		&= {\rm{Tr}}\left( {{\bf{\bar H}}_{\rm{D}}^H{{\left( {{{\bf{H}}_t} + {\mu ^\star}{\bf{I}}} \right)}^{ - 2}}{{{\bf{\bar H}}}_{\rm{D}}}} \right) = {p_{{\rm{D,max}}}}.\notag
	\end{align} 
	Suppose the eigenvalue decomposition ${{\bf{H}}_t} = {{\bf{U}}^H}\Lambda {\bf{U}}$, then 
	\begin{align} \label{bisection}
		f\left( {{\mu ^\star}} \right) &= {\rm{Tr}}\left( {{{\left( {{\bf{W}}_t^\star\left( {{\mu ^\star}} \right)} \right)}^H}{\bf{W}}_t^\star\left( {{\mu ^\star}} \right)} \right) \notag \\
		& = \sum\limits_{i = 1}^{{N_t}} {\frac{{{{\left[ {{\bf{U}}{\bf{\bar H}}_{\rm{D}}{{{\bf{\bar H}}}_{\rm{D}}}^H{{\bf{U}}^H}} \right]}_{\left( {i,i} \right)}}}}{{{{\left( {{{\left[ \Lambda  \right]}_{\left( {i,i} \right)}} + {\mu ^\star}} \right)}^2}}}}  = {p_{{\rm{D,max}}}},
	\end{align}
	which is a monotonically decreasing function. In this case, an optimal $\mu^\star$ can be found via a bisection method by solving (\ref{bisection}) and ${{\bf{W}}_t^{\star}}$ can correspondingly obtained by (\ref{KKT}).
	\vspace{-0.1cm}
	\subsection{Sub-problem 2: Optimizing ${\bf{W}}_r$}
	With fixed ${{\bf{W}}_t}$, ${{\bf{P}}_{\rm{U}}}$, ${\bf{\tilde t}}$, ${\bf{\tilde r}}$, ${\bf{y}}$, and $\boldsymbol{\gamma}$, optimizing ${{\bf{w}}_r}$ in (P3) is to operate (P5)
	\begin{equation}
		\mathop {\max }\limits_{{{\bf{W}}_r}} \  2{\mathop{\rm Re}\nolimits} \left\{ {{\rm{Tr}}\left( {{\bf{Y}}_{\rm{U}}^H{\bf{\bar H}}_{\rm{U}}^H{{\bf{W}}_r}} \right)} \right\} - {\rm{Tr}}\left( {{{\bf{Y}}_{\rm{U}}}{\bf{W}}_r^H{{\bf{H}}_r}{{\bf{W}}_r}{\bf{Y}}_{\rm{U}}^H} \right), \label{sp2}
	\end{equation}
	in which, ${{\bf{H}}_r} = {{\bf{H}}_{\rm{U}}}{{\bf{P}}_{\rm{U}}}{\bf{H}}_{\rm{U}}^H + {{\bf{H}}_{{\rm{SI}}}}{{\bf{W}}_t}{\bf{W}}_t^H{\bf{H}}_{{\rm{SI}}}^H + \sigma _{\rm{U}}^2{\bf{I}}$, ${{\bf{A}}_{\rm{U}}} = {\rm{diag}}\left( {\left\{ {\sqrt {{a_i}\left( {1 + {\gamma _i}} \right)} } \right\}_{i = {K_{\rm{D}}} + 1}^{{K_{\rm{D}}} + {K_{\rm{U}}}}} \right)$, and ${{{\bf{\bar H}}}_{\rm{U}}} = {{\bf{H}}_{\rm{U}}}{{\bf{A}}_{\rm{U}}}{\bf{P}}_{\rm{U}}^{1/2}$. ${{\bf{M}}^{1/2}}$ here is an operation to calculate the element-wise square root of a matrix. Similarly, ${{\bf{H}}_r}$ is a positive semi-definite matrix and thus, (P5) is convex on ${{\bf{W}}_r}$. Moreover, there are no constraints on ${{\bf{W}}_r}$ here because (\ref{1e}) has been eliminated temporarily according to the Proposition 1. By completing the square, (\ref{sp2}) is reformulated as
	\begin{align}\label{15}
		{\bf{W}}_r^\star&=\mathop {{\rm{argmax}}}\limits_{{{\bf{W}}_r}} {\rm{Tr}}\left( {{\bf{\bar H}}_{\rm{U}}^H{\bf{H}}_r^{ - 1}{{{\bf{\bar H}}}_{\rm{U}}}} \right)- \notag \\
		&{\rm{Tr}}\left[ {{{\left( {{{\bf{W}}_r}{\bf{Y}}_{\rm{U}}^H - {\bf{H}}_r^{ - 1}{{{\bf{\bar H}}}_{\rm{U}}}} \right)}^H}{{\bf{H}}_r}\left( {{{\bf{W}}_r}{\bf{Y}}_{\rm{U}}^H - {\bf{H}}_r^{ - 1}{{{\bf{\bar H}}}_{\rm{U}}}} \right)} \right].
	\end{align}
	We can directly obtain the closed-form optimal solution ${\bf{W}}_r^\star = {\bf{H}}_r^{ - 1}{{{\bf{\bar H}}}_{\rm{U}}}{\bf{Y}}_{\rm{U}}^{ - H}$ for (P5) with the optimal value ${\rm{Tr}}\left( {{\bf{\bar H}}_{\rm{U}}^H{\bf{H}}_r^{ - 1}{{{\bf{\bar H}}}_{\rm{U}}}} \right)$. 
	\subsection{Sub-problem 3: Optimizing ${{\bf{P}}_{\rm{U}}}$}
	In the original problem, UL transmitted data interact through MUI, causing power allocation to be coupled in the achievable rate as ratios. However, the introduced two transforms ultimately express the UL transmit power independently for each UL user in the objective function. Moreover, each UL device has its own power constraints, ranging from $0$ to ${p_{{\rm{U,max}}}}$, though we assume a uniform maximum power restriction in this paper. As a result, the sub-problem optimizing ${{\bf{P}}_{\rm{U}}}$ can be solved individually w.r.t. each UL user. Without loss of generality, we take the ${k_{\rm{U}}}$-th UL user as an example
	\begin{align}
		\left( {\rm{P6}}-{k_{\rm{U}}}\right)  \quad \mathop {\max }\limits_{p_{{\rm{U}},{k_{\rm{U}}}}} \quad & {c_{1,{k_{\rm{U}}}}}\sqrt {{p_{{\rm{U}},{k_{\rm{U}}}}}}  - {c_{2,{k_{\rm{U}}}}}{p_{{\rm{U}},{k_{\rm{U}}}}} \label{sp3} \\
		\text{s.t.} \quad &0 \le {p_{{\rm{U}},{k_{\rm{U}}}}} \le {p_{{\rm{U,max}}}},  \notag
	\end{align}
	where
	\begin{subequations}
		\begin{align}
			&{c_{1,{k_{\rm{U}}}}} = 2\sqrt {{a_{{K_{\rm{D}}} + {k_{\rm{U}}}}}\left( {1 + {\gamma _{{K_{\rm{D}}} + {k_{\rm{U}}}}}} \right)} \cdot \notag \\
			&\hspace{10em} {\mathop{\rm Re}\nolimits} \left\{ {y_{{K_{\rm{D}}} + {k_{\rm{U}}}}^*{\bf{h}}_{{\rm{U}},{k_{\rm{U}}}}^H\left( {{\bf{\tilde r}}} \right){{\bf{w}}_{r,{k_{\rm{U}}}}}} \right\},  \notag\\
			&{c_{2,{k_{\rm{U}}}}}={\bf{h}}_{{\rm{IUI}},{k_{\rm{U}}}}^H{\bf{Y}}_{\rm{D}}^H{{\bf{Y}}_{\rm{D}}}{{\bf{h}}_{{\rm{IUI}},{k_{\rm{U}}}}} \notag \\
			&\hspace{10em} +{\bf{h}}_{{\rm{U}},{k_{\rm{U}}}}^H{{\bf{W}}_r}{\bf{Y}}_{\rm{U}}^H{{\bf{Y}}_{\rm{U}}}{\bf{W}}_r^H{{\bf{h}}_{{\rm{U}},{k_{\rm{U}}}}}. \notag
		\end{align}
	\end{subequations}
	As a combination of a square root and an affine function, the convexity of the objective function of (${\rm{P6}}-{k_{\rm{U}}}$) depends on the sign of ${c_{1,{k_{\rm{U}}}}}$. ${c_{2,{k_{\rm{U}}}}}$ is in a positive semi-definite quadratic form, while the sign of ${c_{1,{k_{\rm{U}}}}}$ remains uncertain. If ${c_{1,{k_{\rm{U}}}}} > 0$, the objective function is convex on ${p_{{\rm{U}},{k_{\rm{U}}}}}$ given the box constraint. In this case, the global optimum is achieve at the stationary point (SP) or boundary of the feasible region. The SP is obtained at ${p_{{\rm{U}},{k_{\rm{U}}}}} = \frac{{c_{1,{k_{\rm{U}}}}^2}}{{4c_{2,{k_{\rm{U}}}}^2}} \ge 0$. In the condition $\frac{{c_{1,{k_{\rm{U}}}}^2}}{{4c_{2,{k_{\rm{U}}}}^2}}$ falls within the range $[0,{p_{{\rm{U,max}}}}]$, this point is exactly the optimal power for the ${k_{\rm{U}}}$th UL user. Otherwise, ${p_{{\rm{U}},{k_{\rm{U}}}}^\star} = {p_{{\rm{U,max}}}}$. On the other hand, if ${c_{1,{k_{\rm{U}}}}} \le 0$, the objective function would be a monotonically decreasing function of ${p_{{\rm{U}},{k_{\rm{U}}}}}$ and hence ${p_{{\rm{U}},{k_{\rm{U}}}}^\star}$ should equal to $0$, which means to mute the transmission of this user. While this result may initially seem counter-intuitive—since it suggests that certain users should transmit at less than their maximum power when maximizing data rate—it is, in fact, a rational outcome due to the presence of inter-user interference at both the DL users and the BS. By reducing transmitted power of certain users, overall channel conditions are improved, resulting in enhanced overall system performance.
	\vspace{-0.5cm}
	\subsection{Sub-problem 4: Optimizing ${\bf{\tilde t}}$ and ${\bf{\tilde r}}$} \label{subsp4}
	With given ${{\bf{W}}_t}$, ${{\bf{W}}_r}$, ${{\bf{P}}_{\rm{U}}}$, ${\bf{\tilde r}}$, ${\bf{y}}$, and $\boldsymbol{\gamma}$, maximizing the weighted-sum rate is equivalent to
	\begin{align}
		\left( {\rm{P7}}\right)  \  {\mathop {\min}\limits_{\bf{\tilde t}}} \ & f_4 = -2{\mathop{\rm Re}\nolimits} \left\{ {{\rm{Tr}}\left( {{{\bf{A}}_{\rm{D}}}{{\bf{Y}}_{\rm{D}}}{\bf{W}}_t^H{{\bf{H}}_{\rm{D}}}\left( {{\bf{\tilde t}}} \right)} \right)} \right\} \notag \\
		&+ {\rm{Tr}}\left( {{\bf{Y}}_{\rm{D}}^H{\bf{H}}_{\rm{D}}^H\left( {{\bf{\tilde t}}} \right){{\bf{W}}_t}{\bf{W}}_t^H{{\bf{H}}_{\rm{D}}}\left( {{\bf{\tilde t}}} \right){{\bf{Y}}_{\rm{D}}}} \right) \notag \\
		& + {\rm{Tr}}\left( {{{\bf{W}}_t}{\bf{W}}_t^H{\bf{H}}_{{\rm{SI}}}^H\left( {{\bf{\tilde t}}} \right){{\bf{W}}_r}{\bf{Y}}_{\rm{U}}^H{{\bf{Y}}_{\rm{U}}}{\bf{W}}_r^H{{\bf{H}}_{{\rm{SI}}}}\left( {{\bf{\tilde t}}} \right)} \right) \label{sp4}  \\
		\text{s.t.} \quad & (\ref{1a}), (\ref{1b}) \notag. 
	\end{align}
	Since we have assumed square reconfigurable regions, (\ref{1a}) remains convex w.r.t. ${\bf{\tilde t}}$. (\ref{sp4}) is arranged in a quadratic form involving ${{{\bf{H}}_{\rm{D}}}\left( {{\bf{\tilde t}}} \right)}$ and ${{{\bf{H}}_{{\rm{SI}}}}\left( {{\bf{\tilde t}}} \right)}$ with positive semi-definite coefficient matrices, however, it remains non-convex on ${\bf{\tilde t}}$ due to the presence of the exponential function. Additionally, (\ref{1b}) is also inherently a non-convex constraint. To efficiently solve (P7), \blue{we propose applying a BSUM \cite{BSUM} framework to tackle the non-convexity of $f_4$. Although both BSUM and SCA follow a similar iterative optimization approach, they differ fundamentally in principle. Specifically, BSUM emphasizes block-wise majorization with upper-bound surrogates, whereas SCA emphasizes approximate convexification of the problem as a whole \cite{BSUM, SCA}. This difference makes BSUM more suitable for the block structure in our problem and achieves a better convergence.} Consequently, instead of solving (P7) directly, we iteratively optimize upper bounds of (\ref{sp4}) in an antenna-by-antenna manner. \\
	\indent In the confined reconfigurable size, (\ref{sp4}) is smooth over ${{\bf{\tilde t}}}$ and differentiable with Lipschitz continuous gradient. \blue{In this case, we adopt a quadratic function to locally upper-bound (\ref{sp4}) within a certain neighbor of a given point since it well balances between the complexity and performance.} Denoting the the optimized positions for the $n_t$-th PRA at the ($m+1$)-th BSUM iteration by ${{{\bf{\tilde t}}}^m} = \left[ {{\bf{t}}_1^{m + 1}, \cdots ,{\bf{t}}_{{n_t} - 1}^{m + 1},{\bf{t}}_{{n_t}}^m, \cdots ,{\bf{t}}_{{N_t}}^m} \right]$, then we define the quadratic surrogate function $u\left( {{{\bf{t}}_{{n_t}}};{{{\bf{\tilde t}}}^m}} \right)$ w.r.t. the $n_t$-th transmit PRA position in the ($m+1$)-th BSUM iteration as (\ref{20}) in the next page. In (\ref{20}),
	\begin{figure*}[t]
		\centering
		\begin{equation} \label{20}
			\begin{aligned}
				&{f_4}\left( {{{\bf{t}}_{{n_t}}};{{{\bf{\tilde t}}}^m}} \right) = {f_4}\left( {{{{\bf{\tilde t}}}^m}} \right) + {\nabla _{{{\bf{t}}_{{n_t}}}}}f _4^T\left( {{{{\bf{\tilde t}}}^m}} \right)\left( {{{\bf{t}}_{{n_t}}} - {\bf{t}}_{{n_t}}^m} \right) + \frac{1}{2}{\left( {{{\bf{t}}_{{n_t}}} - {\bf{t}}_{{n_t}}^m} \right)^T}\nabla _{{{\bf{t}}_{{n_t}}}}^2{f _4}\left( {{{{\bf{\tilde t}}}^m}} \right)\left( {{{\bf{t}}_{{n_t}}} - {\bf{t}}_{{n_t}}^m} \right) + o\left( {\left\| {{{\bf{t}}_{{n_t}}} - {\bf{t}}_{{n_t}}^m} \right\|_2^2} \right)\\
				&\le u\left( {{{\bf{t}}_{{n_t}}};{{{\bf{\tilde t}}}^m}} \right)  \buildrel \Delta \over =  {f_4}\left( {{{{\bf{\tilde t}}}^m}} \right) + {\nabla _{{{\bf{t}}_{{n_t}}}}}f _4^T\left( {{{{\bf{\tilde t}}}^m}} \right)\left( {{{\bf{t}}_{{n_t}}} - {\bf{t}}_{{n_t}}^m} \right) + \frac{1}{2}{\left( {{{\bf{t}}_{{n_t}}} - {\bf{t}}_{{n_t}}^m} \right)^T}{\tau_t^m}{\bf{I}}\left( {{{\bf{t}}_{{n_t}}} - {\bf{t}}_{{n_t}}^m} \right)\\
				&= \underbrace {\frac{{{\tau_t^m}}}{2}{\bf{t}}_{{n_t}}^T{{\bf{t}}_{{n_t}}} + {{\left( {{\nabla _{{{\bf{t}}_{{n_t}}}}}{f _4}\left( {{{{\bf{\tilde t}}}^m}} \right) - {\tau_t^m}{\bf{t}}_{{n_t}}^m} \right)}^T}{{\bf{t}}_{{n_t}}}}_{{u_0}\left( {{{\bf{t}}_{{n_t}}};{{{\bf{\tilde t}}}^m}} \right)} + \underbrace {{f _4}\left( {{{{\bf{\tilde t}}}^m}} \right) - {\nabla _{{{\bf{t}}_{{n_t}}}}}f _4^T\left( {{{{\bf{\tilde t}}}^m}} \right){\bf{t}}_{{n_t}}^m + \frac{{{\tau_t^m}}}{2}{{\left( {{\bf{t}}_{{n_t}}^m} \right)}^T}{\bf{t}}_{{n_t}}^m}_{{u_c}\left( {{{\bf{t}}_{{n_t}}};{{{\bf{\tilde t}}}^m}} \right)},
			\end{aligned}
		\end{equation}
		\hrulefill
		\vspace{-0.3cm}
	\end{figure*}
	$u\left( {x;y} \right)$ means that $u$ is a function of $x$ and is unfolded at $y$, ${\nabla _{{{\bf{t}}_{{n_t}}}}}{f _4}\left( {{{{\bf{\tilde t}}}^m}} \right) \in \mathbb{R}^2$ and $\nabla _{{{\bf{t}}_{{n_t}}}}^2{f _4}\left( {{{{\bf{\tilde t}}}^m}} \right) \in \mathbb{R}^{2 \times 2}$ are the gradient vector and Hessian matrix of $f _4$ over ${{\bf{t}}_{{n_t}}}$, respectively, and ${\tau_t^m}$ is a positive number with ${\tau_t^m}{\bf{I}} \succeq \nabla _{{{\bf{t}}_{{n_t}}}}^2{f _4}\left( {{{{\bf{\tilde t}}}^m}} \right)$. A practical choice for ${\tau ^m}$ is the largest eigenvalue of the Hessian matrix. The derivation of the gradient vector and Hessian matrix is provided in Appendix A.  \\
	\indent With the upper bound in (\ref{20}), we resort to solve the highly non-convex sub-problem (P7) as follows, denoted by $\left( {\rm{P8}}-m+1, n_t\right)$,
	\begin{align}
		{\mathop {\min}\limits_{ {\bf{t}}_{\left({n_t}\right)}}} \  & \frac{{{\tau_t^m}}}{2}{\bf{t}}_{\left({n_t}\right)}^T{{\bf{t}}_{\left({n_t}\right)}} + {\left( {{\nabla _{{{\bf{t}}_{\left({n_t}\right)}}}}{f _4}\left( {{{{\bf{\tilde t}}}^m}} \right) - {\tau_t^m}{\bf{t}}_{\left({n_t}\right)}^m} \right)^T}{{\bf{t}}_{\left({n_t}\right)}}   \label{21} \\
		\text{s.t.} \  &{{\bf{t}}_{\left({n_t}\right)}} \in {{\cal R}_{ {n_t}}} = \left\{ {{\bf{t}}\left| {{{\left\| {{\bf{t}} - {{\bf{t}}_{\left( i \right) }}} \right\|}_2} \ge {D_{\min }},1 \le i \le {{n_t}}} \right.} \right\}, \tag{\ref{21}{a}} \label{21a} \\
		&(\ref{1a}).\notag
	\end{align}
	It is noteworthy that the ${\bf{t}}_{\left({n_t}\right)}$ here is the $n_t$-th PRA position solved in the ($m+1$)-th BSUM iteration and not necessarily the position of the $n_t$-th PRA. This is because in BSUM, the blocks, i.e., different transmit PRA positions in this paper, may not be arranged in a cyclic rule \footnote{\blue{The cyclic rule means that the variables are updated in a unchanged cyclical way.}} \cite{BSUM}. Although (\ref{21}) is a convex upper bound of the primal problem, the constraint of (P8), especially the feasible movable region for antenna when $n_t > 1$, is still non-convex, leading to the search of the minimum of (\ref{21}) intractable. \\
	\indent Without the constraints in (P8), we acquire the minimum of the objective function at the SP, i.e. ${\bf{t}}_{\left( {{n_t}} \right)}^\star = {\bf{t}}_{\left( {{n_t}} \right)}^m - \frac{1}{{{\tau_t^m}}}{\nabla _{{{\bf{t}}_{\left( {{n_t}} \right)}}}}{f _4}\left( {{{{\bf{\tilde t}}}^m}} \right)$. If ${\bf{t}}_{\left( {{n_t}} \right)}^\star$ satisfies the constraints in (P8), it is exactly the optimal solution. Otherwise, this SP is invalid. \\
	\indent \textit{Theorem 1: If the SP of (P8) is outside its feasible set, the optimal solution ${\bf{t}}_{\left( {{n_t}} \right)}^\star$ lies on the boundary $\cal B$ of the feasible set ${{\cal R}_{ {n_t}}}$.} \\
	\indent \textit{Proof:} See Appendix B. $\hfill\blacksquare$ \\
	\indent Existing studies predominantly determine the optimal antenna placement using iterative search algorithms such as GD and PSO. However, these methods are highly sensitive to initialization and blocked by disjoint feasible set as illustrated in \cite{HYC1}, often leading to locally optimal solutions or slow convergence. By leveraging the structural properties of (P8) and the geometry of the feasible set, Theorem 1 effectively narrows the search space for ${\bf{t}}_{\left( {{n_t}} \right)}^\star$. \\
	\indent We will provide a detailed procedure for determining ${\bf{t}}_{\left( {{n_t}} \right)}^\star$ when the SP lies within the reconfigurable region, as outlined in Algorithm \ref{alg1}. First, we identify the antennas located within a ${D_{\min }}$-radius from the SP as $\cal A$. If $\cal A$ is empty, the SP itself is the optimal solution. When $\left| \cal A \right| \ne 0$, the optimal solution would lie in the intersections of the antenna’s ${D_{\min }}$-circle with the straight line connecting the antenna and the SP (SCI) or the intersections of these antenna’s circles (CCI). Collecting all calculated points as $\cal P$, the one in the feasible set and nearest to SP is the potential optimal solution. Next, we replace SP by this point and repeat the procedure until the current nearest point satisfies (\ref{21a}), which is the optimal solution. \blue{At the very beginning, the antenna positions are initialized randomly in the feasible space.} Appendix B provides an example for the case $n_t=3$ for this algorithm. The detailed calculation of SCIs and CCIs can be referred to \cite{HYC1} (\textit{Lemma 1 and 2}). Note that the procedure for searching the optimal position proposed here is similar to the one in \cite{HYC1}. However, in \cite{HYC1}, the authors search the optimal point based on a penalty auxiliary variables while our search is based on the stationary point in BSUM iterations.\\
	\begin{algorithm}[tbp]
		\setlength{\textfloatsep}{0.cm}
		\setlength{\floatsep}{0.cm}
		\small
		\caption{Algorithm for finding the optimal position of (P8-$m+1,n_t$)}
		\renewcommand{\algorithmicrequire}{\textbf{Input}}
		\renewcommand{\algorithmicensure}{\textbf{Output}}
		\label{alg1}
		\begin{algorithmic}[1]
			\REQUIRE SP ${\bf{t}}_{\left( {{n_t}} \right)}^{\rm{sp}}$, optimized positions $\mathcal{S} = \left\{ {{\bf{t}}_{\left( i \right)}^m} \right\}_{i = 1}^{{n_t} - 1} \cup \left\{ {\bf{t}}_{\left( i \right)}^{m-1} \right\}_{i = n_t+1}^{{N_t}}$, and minimal inter-antenna distance ${D_{\min }}$.
			\ENSURE Optimal position for the $n_t$th antenna ${\bf{t}}_{\left( {{n_t}} \right)}^\star$.
			\STATE Initialize the center point ${\bf{t}}_c$ by ${\bf{t}}_{\left( {{n_t}} \right)}^{\rm{sp}}$;
			\STATE Find the antennas $\mathcal{A}$ within the ${D_{\min }}$-radius of ${\bf{t}}_c$ from $\mathcal{S}$;
			\IF{$\mathcal{A}$ is empty}
			\STATE ${\bf{t}}_{\left( {{n_t}} \right)}^\star = {\bf{t}}_c$;
			\ELSE
			\FOR {all antennas in $\mathcal{A}$} 
			\STATE Calculate SCIs and CCIs between these antenna and store in $\mathcal{P}$;
			\ENDFOR
			\STATE ${\bf{t}}_{\left( {{n_t}} \right)}^ \star  = \mathop {\arg \min }\limits_{{\bf{t}} \in {\cal P},{{\left\| {{\bf{t}} - {\bf{a}}} \right\|}_2}{ \ge {D_{\min }}},\forall {\bf{a}} \in {\cal A}} {\left\| {{\bf{t}} - {{\bf{t}}_c}} \right\|_2}$;
			\STATE ${\bf{t}}_c = {\bf{t}}_{\left( {{n_t}} \right)}^\star$;
			\ENDIF
			\STATE Repeat step 2 - 13 until the optimal position is acquired at step 4.
		\end{algorithmic}
	\end{algorithm}	
	\indent Overall, we propose a BSUM-based algorithm to solve (P7), where the original non-convex problem is decomposed into a sequence of upper bound minimizations. In each iteration, the positions of $N_t$ antennas are determined following a randomized approach to mitigate the risk of convergence to a local optimum. For each antenna position update, the objective function is upper-bounded using its second-order Taylor expansion, and the optimal position is searched based on geometric properties. The BSUM iterations terminate once a predefined convergence criterion is met. \\
	\indent \blue{Next, we analyze the convergence of the proposed BSUM algorithm. While BSUM is guaranteed to converge to at least a stationary solution under certain conditions \cite{BSUM}, in our problem, the constraints are extremely intractable and deviate from the assumption in \cite{BSUM}. Specifically, the minimum inter-antenna distance constrain forms a disjoint non-convex feasible space, as illustrated in \cite{HYC1}. In the case the stationary solution falls outside the feasible set, the stationary solution cannot be guaranteed. Therefore, we establish the convergence of our algorithm by demonstrating its monotonicity.} We have 
	\begin{equation} \label{22}
		\begin{aligned}
			{f _4}\left( {{{{\bf{\tilde t}}}^m}} \right) &\overset{(\alpha_1)}{=} u\left( {{{{\bf{\tilde t}}}^m};{{{\bf{\tilde t}}}^m}} \right) = {u_0}\left( {{{{\bf{\tilde t}}}^m};{{{\bf{\tilde t}}}^m}} \right) + {u_c}\left( {{{{\bf{\tilde t}}}^m};{{{\bf{\tilde t}}}^m}} \right) \\
			&\overset{(\alpha_2)}{\ge} {u_0}\left( {{\bf{\tilde t}}_{ - 1}^{m + 1};{{{\bf{\tilde t}}}^m}} \right) + {u_c}\left( {{{{\bf{\tilde t}}}^m};{{{\bf{\tilde t}}}^m}} \right) = u\left( {{\bf{\tilde t}}_{ - 1}^{m + 1};{{{\bf{\tilde t}}}^m}} \right)\\
			&\overset{(\alpha_3)}{\ge} {f _4}\left( {{\bf{\tilde t}}_{ - 1}^{m + 1}} \right) = u\left( {{\bf{\tilde t}}_{ - 1}^{m + 1};{\bf{\tilde t}}_{ - 1}^{m + 1}} \right) \\
			&\ \ge {u_0}\left( {{\bf{\tilde t}}_{ - {n_t}}^m;{\bf{\tilde t}}_{ - {n_t} + 1}^{m + 1}} \right) + {u_c}\left( {{\bf{\tilde t}}_{ - {n_t} + 1}^{m + 1};{\bf{\tilde t}}_{ - {n_t} + 1}^{m + 1}} \right) \\
			&\ \ge u\left( {{{{\bf{\tilde t}}}^{m + 1}};{\bf{\tilde t}}_{ - {N_t} + 1}^{m + 1}} \right) \ge {f _4}\left( {{{{\bf{\tilde t}}}^{m + 1}}} \right), \\
		\end{aligned}
	\end{equation}
	where ${\bf{\tilde t}}_{ - i}^{m + 1} = \left[ {{\bf{t}}_{_{\left( 1 \right)}}^{m + 1}, \cdots ,{\bf{t}}_{_{\left( i \right)}}^{m + 1},{\bf{t}}_{_{\left( {i + 1} \right)}}^m, \cdots ,{\bf{t}}_{_{{N_t}}}^m} \right]$. In (\ref{22}), the inequality ($\alpha_1$) holds for the tightness of the upper bound at ${{{\bf{\tilde t}}}^m}$. Then, since (P8) minimizes this upper bound within the feasible set, ($\alpha_2$) holds and the equality holds when the optimized position is identical to the previous one. As demonstrated above, the gradient function of $f_4$ is Lipschitz continuous. Consequently, at ${{\bf{\tilde t}}_{ - 1}^{m + 1}}$, $u\left( { \cdot ;{{{\bf{\tilde t}}}^m}} \right)$ still upper-bounds $f_4$ leading to ($\alpha_3$). These inequalities ensure the descent in the object value in every single antenna position optimization process. Furthermore, after updating $N_t$ positions in one iteration, the value of $f_4$ keeps non-increasing. Recalling that $f_4$ is a convex quadratic function of ${{\bf{g}}\left( {{{\bf{t}}_i}} \right)}$ and ${{{\bf{g}}_{{\rm{SI}}}}\left( {{{\bf{t}}_i}} \right)}$ and meanwhile their elements are all 1-module, it is easy to see $f_4$ would have a lower bound. Overall, $\left\{ {{f_4}\left( {{{{\bf{\tilde t}}}^m}} \right)} \right\}_{m = 1}^\infty$ is a non-increasing sequence with a lower bound and the proposed BSUM algorithm is guaranteed to converge to at least a sub-optimal solution. \\
	\indent The marginal problem for ${\bf{\tilde r}}$ is
	\begin{align} 
		\left( {\rm{P9}}\right)  \  {\mathop {\min}\limits_{\bf{\tilde r}}} \  &f_5 = -2{\mathop{\rm Re}\nolimits} \left\{ {{\rm{Tr}}\left( {{{\bf{A}}_{\rm{U}}}{\bf{P}}_{\rm{U}}^{1/2}{{\bf{Y}}_{\rm{U}}}{\bf{W}}_r^H{{\bf{H}}_{\rm{U}}}\left( {{\bf{\tilde r}}} \right)} \right)} \right\} \notag \\
		&+ {\rm{Tr}}\left( {{{\bf{P}}_{\rm{U}}}{\bf{H}}_{\rm{U}}^H\left( {{\bf{\tilde r}}} \right){{\bf{W}}_r}{\bf{Y}}_{\rm{U}}^H{{\bf{Y}}_{\rm{U}}}{\bf{W}}_r^H{{\bf{H}}_{\rm{U}}}\left( {{\bf{\tilde r}}} \right)} \right) \notag \\ 
		&+ {\rm{Tr}}\left( {{{\bf{W}}_t}{\bf{W}}_t^H{\bf{H}}_{{\rm{SI}}}^H\left( {{\bf{\tilde r}}} \right){{\bf{W}}_r}{\bf{Y}}_{\rm{U}}^H{{\bf{Y}}_{\rm{U}}}{\bf{W}}_r^H{{\bf{H}}_{{\rm{SI}}}}\left( {{\bf{\tilde r}}} \right)} \right)  \notag \\
		\text{s.t.} \  & (\ref{1a}), (\ref{1c}) \notag. 
	\end{align}
	Similar to (P8), (P9) can be solved using the same BSUM-based algorithm as used previously. The analysis of Theorem1, solution for ${\bf{\tilde r}}$, and convergence discussion closely follow the approach used for solving ${\bf{\tilde t}}$. Therefore, for brevity, we omit the detailed discussion of solving (P9) here. \\
	\indent Overall, the FP and BSUM-based AO algorithm for (P1) is summarized in Algorithm \ref{alg2}. \blue{Since the stationary solution is already not guaranteed by the antenna position optimization. We establish the convergence of the AO also by demonstrating its monotonicity and will show the convergence by simulation in Section \ref{sim}. Specifically, in each iteration of Algorithm \ref{alg2}, the objective function is monotonically increasing while remaining bounded by the imposed power constraints, thereby ensuring the convergence of the proposed algorithm.}
	\begin{algorithm}[bp]
		\setlength{\textfloatsep}{0.cm}
		\setlength{\floatsep}{0.cm}
		\small
		\caption{FP-based AO Algorithm for (P1)}
		\renewcommand{\algorithmicrequire}{\textbf{Input}}
		\renewcommand{\algorithmicensure}{\textbf{Output}}
		\label{alg2}
		\begin{algorithmic}[1]
			\REQUIRE User weights $\left\{ {{a_i}} \right\}_{i = 1}^{{K_{\rm{D}}} + {K_{\rm{U}}}}$, CSI (AoD/AoAs, PRM, and noise power).
			\ENSURE Optimal transmit and receive beamformers ${{\bf{W}}_t}$ and $ {{\bf{W}}_r}$, MA positions ${\bf{\tilde t}}$ and ${\bf{\tilde r}}$, and UL user powers ${{\bf{P}}_{\rm{U}}}$.
			\STATE Initialize designed variables to feasible values;
			\REPEAT
			\STATE Update $\bf{y}$ by (\ref{y});
			\STATE Update ${\bm{\gamma}}$ by (\ref{gamma});
			\STATE Update ${\bf{W}}_t$ by (\ref{KKT});
			\STATE Update ${\bf{W}}_r$ by (\ref{15});
			\STATE Update ${{\bf{P}}_{\rm{U}}}$ by (\ref{sp3});
			\REPEAT
			\STATE Update ${\bf{t}}_{\left( {{n_t}} \right)}$ iteratively following Algorithm 1;
			\UNTIL{BSUM converges}
			\REPEAT
			\STATE Update ${\bf{r}}_{\left( {{n_r}} \right)}$ iteratively following Algorithm 1;
			\UNTIL{BSUM converges}
			\UNTIL{AO converges}
		\end{algorithmic}
	\end{algorithm}	
	
	\subsection{Complexity analysis}
	In each iteration of the outer AO algorithm, assuming ${K_{\rm{D}}} = {K_{\rm{U}}} = K$, ${N_t} = {N_r} = N$, and equal path numbers $L$, the computational complexity in optimizing $\mathbf{y}$, $\boldsymbol{\gamma}$, ${{\bf{W}}_r}$, and ${{\bf{P}}_{\rm{U}}}$ are $\mathcal{O} \left( KN \right)$, $\mathcal{O} \left( {K^2N} \right)$, $\mathcal{O} \left( {{N^3} + NK\max \left( {N,K} \right)} \right)$, and $\mathcal{O} \left( {{N^2}K} \right)$, respectively. Because of the bisection method, the complexity for calculating ${{\bf{W}}_t}$ scales with $\mathcal{O} \left( {{K^3}N}{\log _2}d \right)$, where $d$ is the initial range for bisection search. The complexity of proposed BSUM for ${\bf{\tilde t}}$ and ${\bf{\tilde r}}$ is mainly from the calculation of gradient vectors and Hessian matrices, which are all at most $\mathcal{O} \left( {{L^4}{N^2}{K^2}} \right)$. In fact, because there are a lot of common variables, the complexity can be further reduced. \\
	\indent For the proposed Algorithm \ref{alg1}, the computational complexity for finding the optimal antenna positions is $\mathcal{O} \left( N^{\frac{N-1}{\beta}} \right)$, where $\beta$ depends on the condition of problem. This complexity is prohibitive if the number of antennas is large. To address this issue, we introduce a practical modification for the Algorithm \ref{alg1}. In the Step 2, instead of considering all antennas in each iteration, we can exclude antennas that have already been involved in previous rounds. This means that the searching set is gradually shrunk, significantly reducing the complexity to $\mathcal{O} \left( N^2\right) $ even in the worst case while maintaining negligible performance loss, as demonstrated in the simulation results. With this modification, the total computational complexity of proposed Algorithm \ref{alg2} is $\mathcal{O} \left( {{I_f}\left( {{I_b}{L^4}{N^2}{K^2} + {N^3} + NK\max \left( {N,K} \right) + {K^3}N{{\log }_2}d} \right)} \right)$, with the iteration number of FP and BSUM as $I_f$ and $I_b$, respectively.
	
	\section{Simulation Results} \label{sim}
	\subsection{Simulation settings}
	We consider a FD BS simultaneously serving ${K_{\rm{D}}} = {K_{\rm{U}}} = K$ DL and UL users in our simulations. The users are randomly and uniformly distributed within a distance between 20m and 100m to the BS, i.e., $d_i \sim U(20, 100), 1\le i \le {2K}$. At the BS, we assume the transmitter and receiver are equipped with an equal number of PRAs, i.e., $N_t=N_r=N$. Each user-BS channel is assumed to have $L$ resolvable propagation paths, while the SI channel contains $L_{\rm{SI}}$ paths. Under this model, the PRMs are diagonal, and the elevation and azimuth AoD/AoAs for each users are independently and uniformly distributed between 0 and $\pi$. Furthermore, the diagonal elements of the PRMs in both the DL and UL channels are assumed to follow a CSCG distribution with zero mean and variance $\frac{{\rho_0}{d_{i}^{-\alpha}}}{L}$, i.e. ${\cal CN} \left( 0,\frac{{\rho_0}{d_{i}^{-\alpha}}}{L} \right)$, where $\rho_0$ represents the reference path loss at 1m, and $\alpha$ denotes the path loss exponent. For the SI channel, we assume that the elements in $\Sigma_{\rm{SI}}$ follow a CSCG distribution, i.e., ${\cal CN} \left( 0,\frac{\rho_{\rm{SI}}}{L} \right) $. The background noise power at both the DL users and the BS receiver is set equal, i.e., $\sigma _{{\rm{D}}, i}^2=\sigma _{\rm{U}}^2=\sigma^2$. The IUI channel coefficient is modeled as ${{h_{{\rm{IUI}},{k_{\rm{D}}},{k_{\rm{U}}}}}} \sim {\cal CN} \left( 0,\rho_{\rm{IUI}} \right)$. Unless otherwise stated, the key system parameters are listed in Table \ref{tab1}. The weights for all DL and UL users are assumed to be equal, given by $\frac{1}{{{K_{\rm{D}}} + {K_{\rm{U}}}}}$ in this section. All the results are averaged over 10000 independent channel realizations in this section.
	\begin{table}[t]
		\small
		\centering
		\caption{Simulation settings}
		\label{tab1}
		\begin{tabular}{|c|c|c|}
			\hline
			\textbf{Parameter}       & \textbf{Description}                           & \textbf{Value}         \\ \hline
			$K$                        & Number of DL/UL users                             & 4                      \\ \hline
			$N$                        & Number of antennas                             & 4                      \\ \hline
			$A \times A$                        & Size of movable region                         & $4\lambda\times4\lambda$                      \\ \hline
			$D_{min}$                   & Minimum inter-antenna distance                 & $\lambda/2$               \\ \hline
			$L$                        & Path number of DL/UL channels & 8                      \\ \hline
			$L_{\rm{SI}}$                    & Path number of SI channel                      & 6                      \\ \hline
			$\rho_{\rm{SI}}$                  & SI cancellation coefficient                   & -90dB                  \\ \hline
			$\rho_0$                   & Reference path loss                            & -40dB                  \\ \hline
			$\rho_{\rm{IUI}}$                   & IUI coefficient                            & -90dB                  \\ \hline
			$\alpha$                    & Path loss exponent                             & 2.8                    \\ \hline
			$\sigma^2$ & Environmental noise power                      & -90dBm                 \\ \hline
			$p_{\rm{D,max}}$                 & Maximum DL transmit power                      & 40dBm                  \\ \hline
			${p_{{\rm{U,max}}}}$                     & UL transmit power                              & 10dBm                  \\ \hline
			$\epsilon$                  & Convergence threshold                          & $10^{-3}$ \\ \hline
			$f_c$                     & Carrier frequency                              & 30GHz                  \\ \hline
		\end{tabular}
	\end{table}
	
	To validate the performance of the proposed algorithm, referred to as ``\textbf{FP-BSUM-FD}'' in this paper, we compare it against several typical baseline approaches: \\
	\noindent \textbf{FP-GD-FD}: FP is applied for beamforming optimization, but the antenna positions are optimized using GD as in \cite{FPGD}. \\
	\noindent \textbf{FP-SCA-FD}: FP is used for beamforming optimization, while the SCA method in \cite{MIMOMA} is employed for antenna position optimization. Specifically, the non-convex objective function and position constraints are relaxed using first- and second-order Taylor expansions, and the problem is solved via a optimization toolbox. \\
	\noindent \textbf{SCA-BSUM-FD}: Beamforming is directly optimized using SCA as in \cite{MAS-SFD-MU}, while antenna positions are optimized using the BSUM approach proposed in this paper. Note that nature-inspired PSO is not considered in this section. \\
	\noindent \textbf{FPAS-FD}: The antennas are arranged in a uniform planar array (UPA) with a spacing of $\lambda/2$ at both the transmit and receive ends of the BS. The beamforming and power allocation are optimized using the proposed FP-based approach.\\
	\noindent \textbf{FP-BSUM-HD}: The BS, equipped with transmit PRAs, operates in a time-division HD mode to deliver DL service. Variables are optimization by the proposed FP-BSUM approach. 
	\vspace{-0.8cm}
	\subsection{\blue{Computational} and Convergence Evaluation}
	\begin{figure}[t]
		\centering
		\includegraphics[scale=0.5]{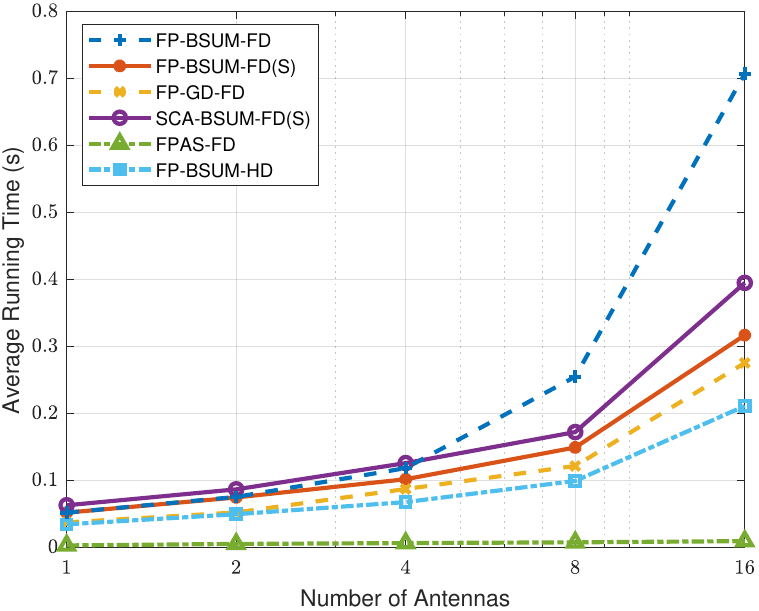}
		\caption{Running time of different algorithms}	
		\label{figrt}
	\end{figure}
	\begin{figure}[t]
		\centering
		\includegraphics[scale=0.5]{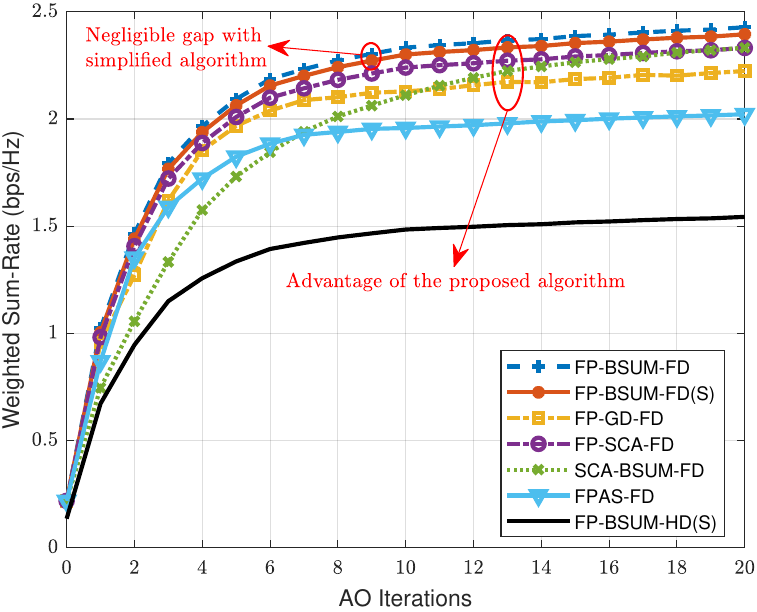}
		\caption{Convergence of different algorithms}	
		\label{figcv}
		\vspace{-0.3cm}
	\end{figure}
	\begin{figure*}[tb]
		\centering
		\subfloat[Number of Antennas]{
			\label{figsuca}
			\includegraphics[scale=0.41]{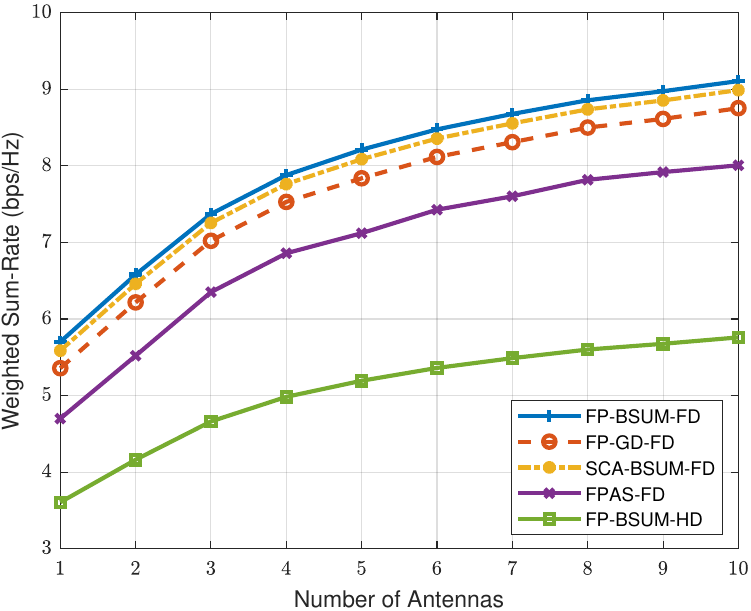}
		}
		\subfloat[Normalized Reconfigurable Region]{
			\label{figsucb}
			\includegraphics[scale=0.41]{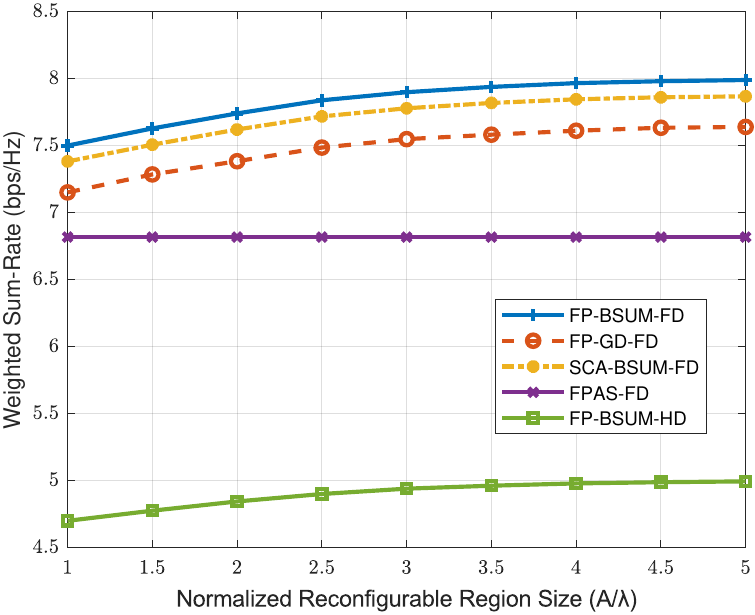}
		}
		\subfloat[DL Transmit Power]{
			\label{figsucc}
			\includegraphics[scale=0.41]{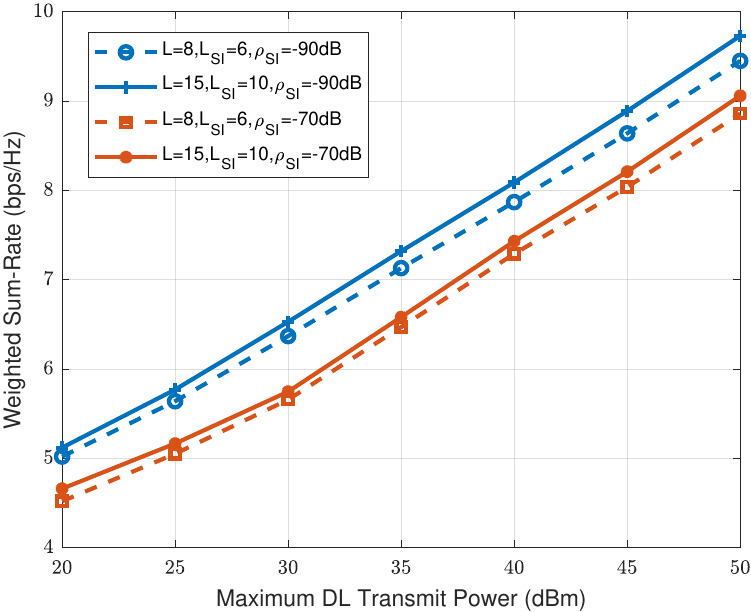}
		}
		\caption{Achievable rates versus different parameters $(K=1)$}	
		\label{figsucom}
	\end{figure*}
	Fig.\ref{figrt} illustrates the average running time per outer iteration of different algorithms, tested on a laptop with an Inter i9-13980HX CPU. In this figure, ``(s)'' refers to the simplified Algorithm \ref{alg1}. For the cases where the number of antennas equals to 1 or 2, the number of users is set equal to the number of antennas. For all other cases, the number of users is fixed at 4. From Fig. \ref{figrt}, we observe that as the number of antennas increases, the computational complexity of all schemes increases accordingly, with Algorithm \ref{alg1} experiencing the most significant growth. However, this complexity can be greatly reduced through the proposed simplification. We do not provide the running time for the ``FP-SCA-FD'' approach, as the execution time when using the optimization toolbox is uncontrollable when a great number of constraints are involved. \\
	\indent In Fig. \ref{figcv}, we further evaluate the convergence behavior of different algorithms under the default environment. The results validate the analysis in Subsection \ref{subsp4} and confirm the convergence of the proposed approach. Notably, the FP-based AO exhibits rapid convergence within 5–10 iterations, whereas the SCA-based outer AO requires approximately 15 iterations to reach a stable performance level. Among the approaches for antenna position optimization, the proposed BSUM achieves the highest weighted sum-rate, albeit at the cost of increased complexity. However, the simplification of Algorithm \ref{alg1} significantly reduces complexity with only marginal performance loss. Thus, for the rest of this section, we use the proposed algorithm with this simplification for simulations.
	\vspace{-0.3cm}
	\subsection{Performance Evaluation under Different Conditions}
	\begin{figure*}[t]
		\centering
		\subfloat[Number of Antennas]{
			\label{figmuca}
			\includegraphics[scale=0.41]{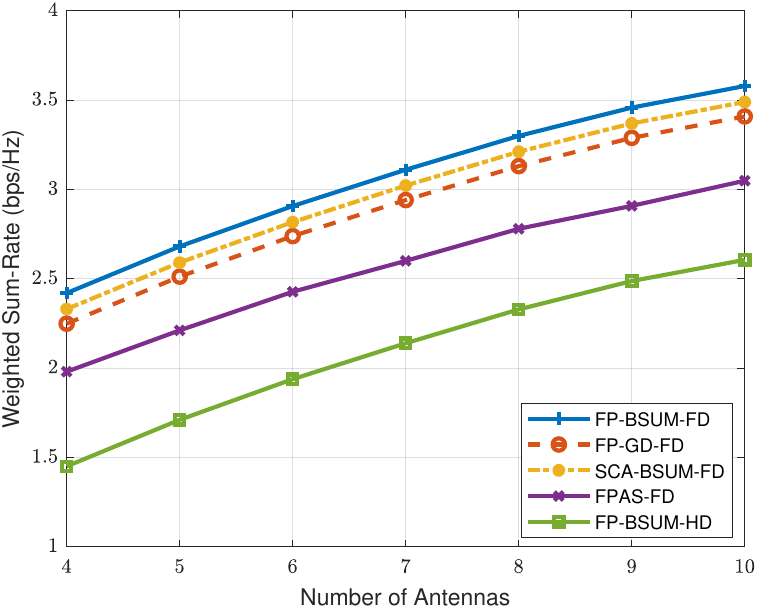}
		}
		\subfloat[Normalized Reconfigurable Region]{
			\label{figmucb}
			\includegraphics[scale=0.41]{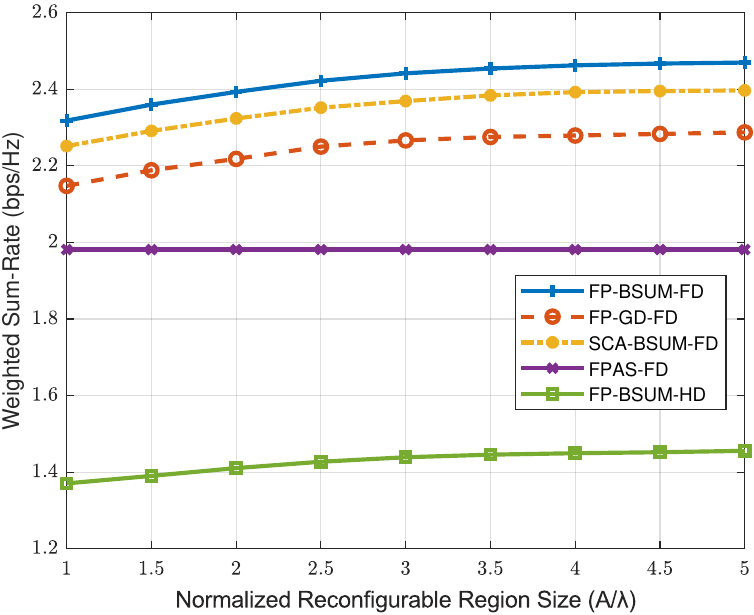}
		}
		\subfloat[DL Transmit Power]{
			\label{figmucc}
			\includegraphics[scale=0.41]{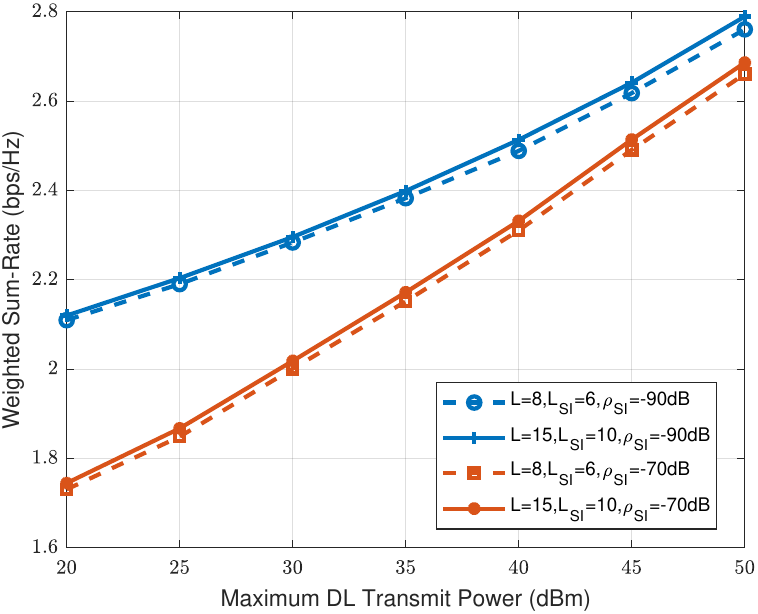}
		}
		\caption{Achievable rates versus different parameters $(K=4)$}	
		\label{figmucom}
	\end{figure*}
	We evaluate the achievable weighted sum-rate of the proposed and baseline algorithms under variations in the number of antennas, reconfigurable region size, and DL transmit power to assess the performance of the proposed approach. The comparison is conducted in two scenarios, serving different number of users, as illustrated in Fig.\ref{figsucom} and Fig.\ref{figmucom}. \\
	\indent In Fig.\ref{figsuca} and Fig.\ref{figmuca}, we show the achievable sum-rate versus the number of antennas. Benefiting from improved spatial multiplexing and diversity, and beamforming gains brought by a larger antenna array, all schemes achieve higher weighted sum-rates. This benefit is remarkable at the beginning but becomes marginal as the number of antennas increases. Furthermore, the results validate the performance advantages of integrating PRA into MU-FD-MIMO systems and the superiority of the proposed algorithm. Specifically, with PRA, the achievable rate increases by more than 20\% with a single antenna. As the number of antennas grows, the improvement of PRAS slightly surpasses that of FPAS, though the gain decreases to approximately 12.5\%. Notably, by comparing the blue and purple lines, we observe that the FD-MIMO system, assisted by PRA, requires only half the number of antennas to achieve a comparable communication rate. These results highlight the advantages of integrating PRAs into the MIMO systems.  \\ 
	\indent We present the achievable sum-rate as a function of the size of normalized reconfigurable region in Fig.\ref{figsucb} and Fig.\ref{figmucb}. The results show that the achievable rate of the PRA-aided system increases as the reconfigurable region expands, however, the growth becomes marginal when the region is sufficiently large. Specifically, the weighted sum-rate improves by 7\% when the size of reconfigurable region increases from $\lambda \times \lambda$ to $3\lambda \times 3\lambda$, whereas further expansion to $5\lambda \times 5\lambda$ results in only a 1\% gain. This trend can be understood intuitively: considering a single-antenna scenario, the performance gain from enlarging the reconfigurable region is proportional to the probability that the peak of the channel amplitude falls outside the current region. Given the periodic nature of the channel field, a larger region is more likely to contain a channel peak, explaining the diminishing gains as the size increases. On the other hand, the achievable rate of FPAS remains unchanged as the reconfigurable region size increases because it lacks the ability to exploit the additional DoF provided by antenna repositioning. \\
	\indent Fig.\ref{figsucc} and Fig.\ref{figmucc} illustrate the performance variation with the change of DL transmit power, number of scattering path, and SI suppression coefficient. The results show that increasing the DL transmit power from 20dBm to 30dBm, 40dBm, and 50dBm leads to a 9\%, 19\%, and 32\% increase in the achievable rate, respectively. This demonstrates that as DL transmit power increases, the achievable rate also grows, and the growth rate accelerates. Moreover, when the environment has more propagation paths, the sum-rate has an increment since a richer scattering environment would potentially provide additional spatial diversity and multiplexing gain, and increase the condition number of channel matrix. Conversely, weaker SI suppression negatively impacts system performance. For instance, if the SI suppression coefficient degrades from -90dB to -70dB, the weighted sum-rate decreases by approximately 5\%–10\% on average.
	\vspace{-0.5cm}
	\subsection{Robustness}
	\begin{figure}[bt]
		\centering
		\includegraphics[scale=0.5]{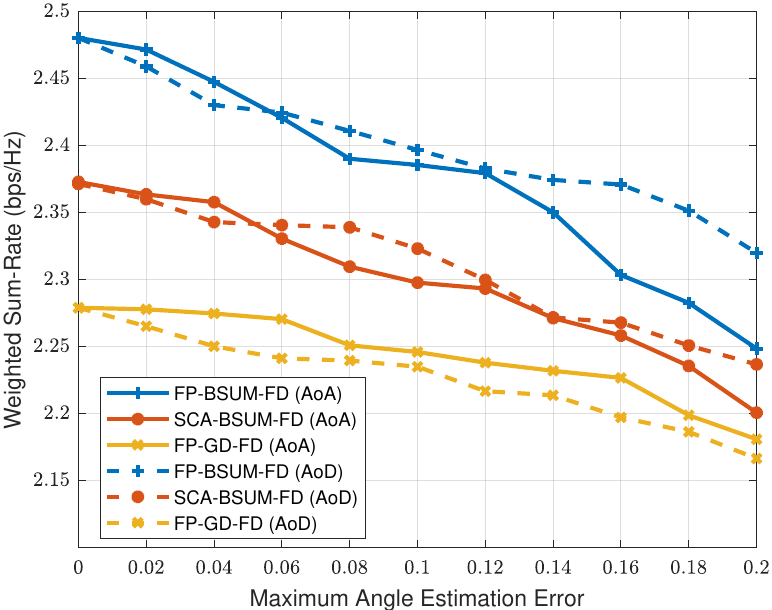}
		\caption{Performance versus angle estimation error}	
		\label{figr1}
	\end{figure}
	\begin{figure}[bt]
		\centering
		\includegraphics[scale=0.5]{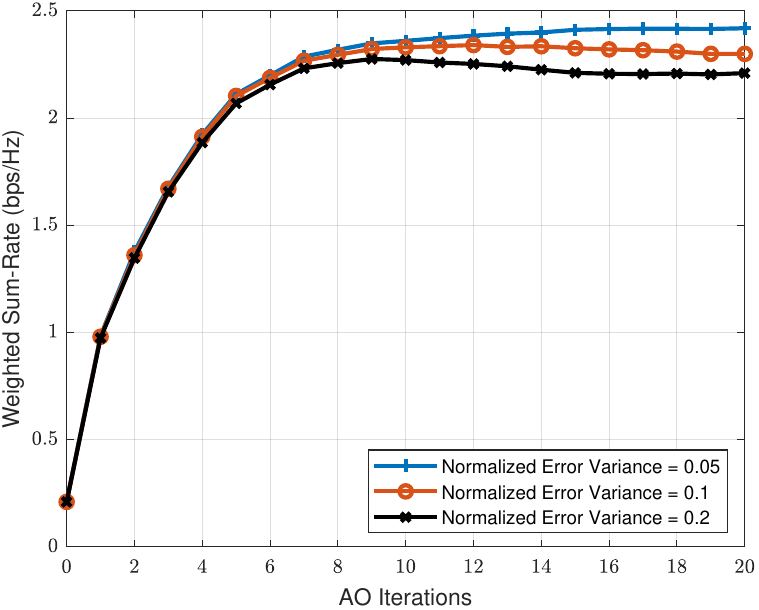}
		\caption{Converging process with errors in PRMs estimation}	
		\label{figr2}
	\end{figure}
	\blue{Hardware implementation and channel estimation are beyond the scope of this paper, and we assume perfect antenna position adjustment and full channel knowledge based on the techniques in \cite{CE-MAS, CE-FAS}. However, in practical scenarios, estimation errors in channel state information and hardware imperfection are inevitable due to background noise, dynamic environments, and hardware constraints. These errors would affect the acquisition of key channel parameters such as angle estimations for AoA/AoD and the acquisition of PRMs. We assess the impact of these imperfections by evaluating the robustness of the proposed algorithm under key channel parameter errors. This analysis provides insights into the algorithm’s resilience in practical deployment scenarios.} \\
	\indent To simulate angle estimation errors, we introduce random noise uniformly distributed between $-\frac{\theta_m}{2}$ and $\frac{\theta_m}{2}$ to the ground truth, where $\theta_m$ represents the maximum estimation error. The error on PRMs is assumed to follow a CSCG distribution, hence, an estimated element $\hat{h}$ in a PRM follows ${\cal CN} \left( h, {\left| h \right|^2} \sigma_e^2 \right)$, where $h$ is the ground truth and $\sigma_e^2$ is the variance of normalized PRM error. \\
	\indent We provide the achievable rate of PRAS-aided MU-FD-MIMO system optimized by different algorithms under different angle estimation errors in Fig.\ref{figr1}. The results indicate that the system performance declines as the estimation error increases, regardless of the optimization algorithm. Specifically, the rate performance of the three compared methods decreases by 10\%, 7\%, and 5\% with a 0.2 maximum angle error. Although the proposed algorithm still outperforms the baselines, it is more negatively impacted by angle estimation errors. This is because the sensitivity to input perturbations is linked to the optimality of the algorithm. As more approximations are introduced, the solution becomes more and more suboptimal, making it less sensitive to input variations. Conversely, highly optimized solutions rely more on precise input data, making them more susceptible to performance degradation due to estimation errors.\\
	\indent It is interesting to observe the converging process of the proposed algorithm when the acquired PRM estimations are imperfect, as illustrated in Fig.\ref{figr2}. In the case of small PRM errors (0.05 normalized variance), the proposed algorithm maintains convergence, albeit with slight performance degradation. However, as the variance of normalized PRM error increases to 0.1 and 0.2, the achievable rate undergoes a rise-and-fall process, accompanied by a larger performance loss. The reason is that imperfect PRM misleads the optimization of system parameters. These results provide valuable insights into how the AO iteration can be controlled in practical scenarios with imperfect CSI.
	
	\section{Conclusion}
	In this paper, we explored the performance advantages that the PRA bring to the MU-FD-MIMO system. We first modeled the PRA-aided MU-FD-MIMO system, where the BS serves multiple DL and UL users with single fixed antenna simultaneously via a transmitter and receiver, both of which are equipped with an array of PRAs. Using the weighted sum-rate as a metric, we developed an AO algorithm for maximizing the achievable rate of the considered system. In this AO framework, FP and BUSM - based approaches were utilized to effectively tackle the non-convexity in the problem. Extensive simulation results demonstrated that the FD paradigm significantly outperforms the HD transmission and the integration of PRA to the FD system further led to a considerable achievable rate gain. Furthermore, compared with existing approaches, the proposed algorithm offers better performance in terms of achievable rate and convergence despite slightly weaker robustness on imperfect CSI.\\
	\indent \blue{It is worth noting that this paper focuses on low-mobility environments with a long coherence time allowing for complex joint optimization. In high-mobility scenarios, where the channel coherence time tends to be shorter, statistical CSI-based optimization methods may be more appropriate to reduce the frequency of acquiring CSI and antenna movement. Moreover, although a simplified antenna searching is proposed in this work, some sub-connected structures and algorithms would further improve the efficiency of the proposed system. These are further directions to be pursed in the future.}
	
	\appendices
	\section{Proof of Theorem 1}
	Firstly, we recast (\ref{21}) by neglecting time index, adding some constants, and completing the square,
	\begin{equation} \label{apb1}
		f\left( {\bf{t}} \right) = \frac{{{\tau_t ^m}}}{2}\left\| {{\bf{t}} - {{\bf{t}}_{{\rm{SP}}}}} \right\|_2^2. \notag
	\end{equation}
	Typically, the feasible set is an $A\lambda \times A\lambda$ square with several circles removed. Picking the center of this square as the origin, we express the feasible set by \\
	\begin{align}
		{{\cal R}_{ {n_t}}} &= \left\{ {{{\left[ {x,y} \right]}^T}\left| {\left| x \right| \le \frac{A}{2}\lambda ,\left| y \right| \le \frac{A}{2}\lambda } \right.} \right\}\backslash \notag \\
		&\hspace{7em}\bigcup\limits_{n = 1}^{{n_t}} {\left\{ {{\bf{t}}\left| {\left\| {{\bf{t}} - {{\bf{t}}_{\left( n \right)}}} \right\|_2^2 < {D_{\min }^2}} \right.} \right\}}. \notag
	\end{align} 
	${{\bf{t}}_{\left( 1 \right)}}$ and ${{\bf{t}}_{\left( 2 \right)}}$ are the antenna positions determined in the previous iteration. \\
	\begin{figure}[b]
		\centering
		\includegraphics[scale=0.65]{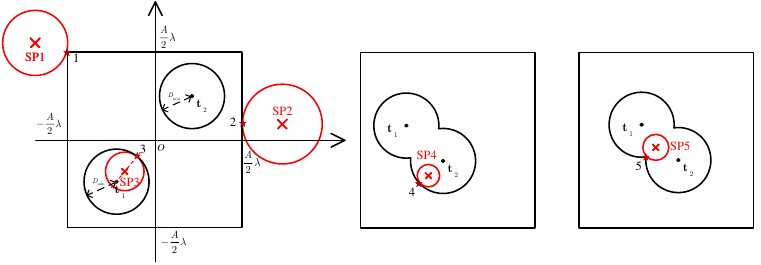}
		\caption{Illustration for different cases}	
		\label{figapb}
	\end{figure}
	\indent In the case that the SP is an exterior point of ${{\cal R}_{ {n_t}}}$, we assume the optimal solution is obtained at an interior point ${\bf{t}}_i$. According to (\ref{apb1}), ${\bf{t}}_i$ is the closest point to ${{\bf{t}}_{{\rm{SP}}}}$ in the feasible set, i.e., $\forall {\bf{t}} \in {{\cal R}_{ {n_t}}},{\left\| {{\bf{t}} - {{\bf{t}}_{{\rm{SP}}}}} \right\|_2} \ge {\left\| {{{\bf{t}}_i} - {{\bf{t}}_{{\rm{SP}}}}} \right\|_2}$. However, connecting ${\bf{t}}_i$ and ${{\bf{t}}_{{\rm{SP}}}}$, the intersection with $\cal B$ will be closer to ${{\bf{t}}_{{\rm{SP}}}}$ than ${\bf{t}}_i$, which is contradictory to the assumption. Consequently, if the SP is an exterior point of the feasible set, the optimal solution must be in $\cal B$. \\
	\indent We take the $n_t=3$ scenario as an example here, as illustrated in Fig. \ref{figapb}. \\
	\noindent \textit{Case 1: SP is outside the square region} \\
	\noindent The red solid cross and line are the SP and contour of (\ref{apb1}), respectively. Shown by SP1 and SP2, the optimal location of case 1 lies on the vertex or the edge of the square. \\
	\noindent \textit{Case 2: SP falls in only one antenna's $D_{min}$-circle} \\
	\noindent Illustrated by SP3 and SP4, ${\bf{t}^\star}$ is the intersection point of this circle and the straight line connecting this antenna and SP (SCI). \\
	\noindent \textit{Case 3: SP falls in two antennas' $D_{min}$-circles} \\
	\noindent In this case, the SCIs w.r.t. both two antennas would fall in the other antennas' circle. To this end, ${\bf{t}^\star}$ is one of the intersection points of these two circle (CCI), as shown in the SP5. \\
	\noindent In the circumstance the edge/vertex point of Case 1 falls in any circle, it follows a similar discussion like Case 2 and 3.
	
	\bibliographystyle{IEEEtran}
	\bibliography{ref_new}
\end{document}